\documentclass[fleqn,10pt]{SelfArx} 

\usepackage[english]{babel} 

\usepackage{lipsum} 

\definecolor{color1}{RGB}{0,0,90} 
\definecolor{color2}{RGB}{0,20,20} 

\usepackage{hyperref} 

\hypersetup{
	hidelinks,
	colorlinks,
	breaklinks=true,
	urlcolor=color2,
	citecolor=color1,
	linkcolor=color1,
	bookmarksopen=false,
	pdftitle={Title},
	pdfauthor={Author},
}

\usepackage{listings}
\usepackage{multirow}
\usepackage[most]{tcolorbox}
\newtcolorbox{notebox}[1][]{
  enhanced,
  breakable,
  title=Note,
  fonttitle=\bfseries,
  coltitle=black,
  colback=blue!3,        
  colframe=blue!50!black,
  boxrule=0.6pt,
  arc=2mm,
  left=1.5mm,right=1.5mm,top=1mm,bottom=1mm,
  #1
}

\usepackage{booktabs}
\usepackage{siunitx}

\newtcolorbox{summarybox}[1][]{
  enhanced,
  breakable,
  title=Summary,
  fonttitle=\bfseries,
  coltitle=black,
  colback=gray!5,
  colframe=gray!60,
  boxrule=0.5pt,
  arc=2mm,
  left=2mm, right=2mm, top=1mm, bottom=1mm,
  #1
}

\usepackage{stfloats}

\PaperTitle{PINNs for Electromagnetic Wave Propagation} 

\Authors{\textbf{Nilufer K. Bulut}, \textit{Izmir, Turkiye}} 
\Correspondence{Correspondence: niluferkbulut@outlook.com}
\affiliation{ORCID: 0009-0005-1015-9570 \quad https://archnilux.github.io} 

\newcommand{\keywordname}{Keywords} 

\Abstract{Physics-Informed Neural Networks (PINNs) solve physical systems by incorporating governing partial differential equations directly into neural network training. In electromagnetism, where well-established methodologies such as FDTD and FEM already exist, new methodologies are expected to provide clear advantages to be accepted. Despite their mesh-free nature and applicability to inverse problems, PINNs can exhibit deficiencies in accuracy and energy metrics compared to FDTD. This study demonstrates that hybrid training strategies can bring PINNs closer to FDTD-level accuracy and energy consistency.

A hybrid methodology addressing common challenges in wave propagation is presented. Causality collapse in time-dependent PINN training is addressed via time marching and causality-aware weighting. To mitigate discontinuities introduced by time marching, a two stage interface continuity loss is applied. To suppress cumulative energy drift in electromagnetic waves, a local Poynting-based regularizer is developed.

In the developed PINN model, high field accuracy is achieved with an average 0.09\% NRMSE and 1.01\% $L^2$ error over time. Energy conservation is achieved with only a 0.024\% relative energy mismatch in the 2D PEC cavity scenario. Training is performed without labeled field data, using only physics-based residual losses; FDTD is used solely for post-training evaluation. The results demonstrate that PINNs can achieve competitive results with FDTD in canonical electromagnetic examples and are a viable alternative.}

\begin{document}

\maketitle 
\tableofcontents 

\thispagestyle{empty} 


\section*{Introduction} 

\addcontentsline{toc}{section}{Introduction} 

Maxwell’s equations are a four-part system of differential equations that forms the cornerstone of classical electromagnetism and describes the space-time dynamics of macroscopic electromagnetic fields. This system of equations reveals the fundamental relationships among the electric field vector E, the magnetic field vector H, the electric flux density D, the magnetic flux density B and the current density J. The electromagnetic properties of the medium are defined by the parameters electric permittivity ($\varepsilon$), magnetic permeability ($\mu$) and electrical conductivity ($\sigma$). The differential-form expressions of the equations are as follows.

\begin{align*} \nabla \times \mathbf{E} &= -\frac{\partial \mathbf{B}}{\partial t} && \text{(Faraday's Law)} \\ \nabla \times \mathbf{H} &= \mathbf{J} + \frac{\partial \mathbf{D}}{\partial t} && \text{(Ampere's Law)} \\ \nabla \cdot \mathbf{D} &= \rho_v && \text{(Gauss' Law)} \\ \nabla \cdot \mathbf{B} &= 0 && \text{(Gauss' Law for Magnetism)} \\ \end{align*}

Different numerical techniques such as the finite difference time domain (FDTD) \cite{1138693}, finite element method (FEM)\cite{CLOUGH199089} and method of moments (MoM) \cite{1447363} have been developed for modeling and analyzing electromagnetic fields. These conventional approaches are based on the discretization of Maxwell's equations and iterative solution algorithms. Therefore, they require intensive computational resources in problems involving complex geometries and variable material properties.

Physics-Informed Neural Networks (PINNs) were introduced between 2017 and 2019 by Maziar Raissi, Paris Perdikaris and George E. Karniadakis \cite{RAISSI2019686}. The PINN method involves directly incorporating physical laws defined by partial differential equations into the neural network training process. Although it can be formulated within a supervised learning framework, it generally reduces dependence on labeled data since it is performed through optimization residuals.

In the literature, adaptations of PINN applications to the time-dependent Maxwell equations have remained limited due to issues such as spectral bias, the sensitivity of electromagnetic problems to scaling, causality violations and loss balancing. The aim of this study is to present a hybrid methodology that addresses the common issues observed in PINN applications for the time-dependent Maxwell equations.

In section 1, a well-posed initial-boundary value problem formulation is obtained for the TM$_z$ mode of the time-dependent Maxwell equations. The non-dimensionalization steps are carried out and the main physics-informed loss components used for the resulting coupled PDE system are defined.

Section 2 focuses on implementation strategies for the PINN architecture established in the previous theoretical section: the neural network structure, the time-marching approach and the sequential training pipeline, the causality-aware weighting scheme, the interface continuity loss that ensures continuity at window interfaces and the Poynting-based regularization that controls the energy dynamics are presented in this section.

In section 3, the metrics used in the study are introduced; the performance of the PINN model is evaluated through field-based error metrics and energy metrics. In addition, three different variants of the Poynting regularizer developed in this study are examined under the ablation studies section.


\section{Theoretical Framework}
\subsection{Well-Posedness}

In their most general form, Maxwell’s equations form a coupled system of partial differential equations. Without appropriate initial and boundary data, the problem is underdetermined and admits infinitely many solutions. Neural networks rely on the universal approximation theorem and can essentially be viewed as function approximators that perform function-space mapping. \cite{HORNIK1989359}

For this reason, in order to solve Maxwell’s equations with PINN architectures, the problem that the model will learn must first be formulated in a well-posed manner. According to Hadamard’s classical approach, three criteria must be satisfied for well posedness: (1) existence of a solution, (2) uniqueness of the solution, and (3) continuous dependence of the solution on the data. \cite{hadamard2014lectures}

In this study, the 2D time-domain Maxwell equations are considered in the $TM_z$ mode. A rectangular cavity with PEC walls is used, and a Gaussian pulse initialization is applied to excite the modal content. Therefore, under the source-free and lossless assumption ($J=0, \sigma = 0$), and together with appropriate initial and boundary conditions, the resulting system defines a well-posed initial–boundary value problem:

\begin{align}
\frac{\partial H_y}{\partial x} - \frac{\partial H_x}{\partial y} - \varepsilon \frac{\partial E_z}{\partial t} &= 0 \nonumber \\
\frac{\partial E_z}{\partial y} + \mu \frac{\partial H_x}{\partial t} &= 0 \nonumber \\
\frac{\partial E_z}{\partial x} - \mu \frac{\partial H_y}{\partial t} &= 0
\end{align}

Modal decomposition is a common strategy in PINN formulations for electromagnetic problems, as it facilitates mode-specific learning and improves numerical conditioning. \cite{pugin2024maxwellpinn} The process of obtaining the well-posed system of equations and the modal decomposition step are described in detail in Appendix A.

For numerical consistency, non-dimensionalization is applied to the well-posed equations. Details of the non-dimensionalization step are provided in Appendix B. The resulting final system of equations is given as follows:

\begin{align} 
\frac{\partial \bar{H_y}}{\partial \bar{x}} - \frac{\partial \bar{H_x}}{\partial \bar{y}} - \frac{\partial \bar{E_z}}{\partial \bar{t}} &= 0 \nonumber \\[0.5em] 
\frac{\partial \bar{E_z}}{\partial \bar{y}} + \frac{\partial \bar{H_x}}{\partial \bar{t}} &= 0 \nonumber \\[0.5em] 
\frac{\partial \bar{E_z}}{\partial \bar{x}} - \frac{\partial \bar{H_y}}{\partial \bar{t}} &= 0 
\end{align}

In the remainder of this work, for notational simplicity, the bar notation will be dropped, but all variables will be implicitly assumed to be non-dimensional.

\subsection{Physics-Informed Loss Components}
For electromagnetic cavity resonator modeling, the loss function formulation consists of partial differential equation (PDE), initial condition (IC) and boundary condition (BC) components. The PDE loss component is defined through residual functions that enforce Maxwell’s curl laws. The boundary conditions loss, which varies depending on the specific scenario, enforces whether the equations satisfy the physical laws at the boundaries. The initial condition loss regularizes the transient electromagnetic response inside the cavity according to the chosen initialization type. In this section, the theoretical development of the loss components is addressed.

\subsubsection*{PDE Loss}
The non-dimensionalized coupled system obtained in Appendix A and Appendix B forms the basis of the PDE loss component. Within the PINN framework, these equations can be reformulated as residual functions that measure to what extent the neural network predictions violate the physical laws:

\begin{align} R_1(x,y,t) &= \frac{\partial H_y}{\partial x} - \frac{\partial H_x}{\partial y} - \frac{\partial E_z}{\partial t} \nonumber \\ 
R_2(x,y,t) &= \frac{\partial E_z}{\partial y} + \frac{\partial H_x}{\partial t} \nonumber \\ 
R_3(x,y,t) &= \frac{\partial E_z}{\partial x} - \frac{\partial H_y}{\partial t} \end{align}

Specifically, $R_1$ enforces Ampere’s law by relating the curl of the magnetic field to the time derivative of the electric field, while $R_2$ and $R_3$ enforce the components of Faraday’s law by coupling the electric field gradients to the time derivatives of the magnetic field.

The PDE loss component is constructed by evaluating these residuals at collocation points distributed over the spatio-temporal domain. For aggregating residual violations, the mean squared error (MSE) metric is used:
\begin{align}
    {L}_{\text{PDE}}(\theta) = \lambda_1\text{MSE}(R_1) + \lambda_2\text{MSE}(R_2) + \lambda_3\text{MSE}(R_3)
\end{align}

Assuming that the residuals ideally vanish, the MSE terms are computed as
\begin{align}
    {MSE}(R_k) = \frac{1}{N_c} \sum_{i=1}^{N_c} |R_k(x_i, y_i, t_i)|^2,\quad k \in {1,2,3}
\end{align}

\subsubsection*{Boundary Condition Loss}
Since a PEC cavity is considered in this study, the boundary condition requires the tangential component of the electric field to vanish on the conductor surface. For a rectangular cavity $(0 \leq X \leq a, 0 \leq Y \leq b)$;

\medskip

($x = X/a$), ($y = Y/a$), and ($\beta = b/a$) are defined. Thus, the cavity boundaries are drawn as $(0 \leq x \leq 1, 0 \leq y \leq \beta)$. The coefficient $\beta$ is included as an aspect ratio term so that, even though the cavity is defined as rectangular, the equality $L=a$ defined during the non-dimensionalization step is preserved.

The cavity walls extend to infinity along the $z$-axis, therefore the $z$ component is tangential along all walls, and a single scalar condition for the electric field must be satisfied on all four walls:
\begin{align*}E_z(0,y,t) = 0 \\
E_z(1,y,t) = 0 \\
E_z(x,0,t) = 0 \\
E_z(x,\beta,t) = 0\end{align*}

Thus, the BC loss can be written as

\begin{align}{L}_{BC}(\theta) = \frac{1}{N_{bc}} \sum_{k=1}^{N_{bc}} |E_z(x_k, y_k, t_k)|^2 \end{align}

In the $TM_z$ mode, imposing $E_z=0$ on PEC walls provides the required electromagnetic boundary condition. The magnetic-field components are then determined implicitly through the PDE residuals and the enforced electric boundary; adding extra boundary conditions on $H_x$ or $H_y$ would overconstrain the system and is therefore avoided.

\subsubsection*{Initial Condition Loss}
In this study, a Gaussian pulse is selected as the excitation. The initial field configuration at $t=0$ for the electric field distribution is expressed as

\begin{align}E_z(x,y,0) = A \cdot \exp(-\frac{(x-x_0)^2 + (y-y_0)^2}{2\sigma^2})\end{align}

The magnetic field components are initialized to zero at $t=0$:
\begin{align*} H_x(x,y,0) = 0 \\ 
H_y(x,y,0) = 0 \end{align*}

The pulse parameters are defined as centered in the cavity at $(x_0, y_0) = (0.5, 0.5 \beta)$, with width $\sigma = 0.1$ and amplitude (A = 1.0). With these parameters, the target initial condition is explicitly reduced to the following form:
\begin{align}E_{z_{tar}}(x,y,0) = A \cdot \exp(-\frac{(x-0.5)^2 + (y-0.5 \beta)^2}{2 \cdot (0.1)^2})\end{align}

The initial condition loss enforces the predicted field distribution at $t=0$:
\begin{align} L_{IC}(\theta)=&\frac{1}{N_{ic}}[ \sum_{j=1}^{N_{ic}}\bigl|E_z(x_j,y_j,0;\theta)-E_{z_{tar}}(x_j,y_j,0)\bigr|^2 \nonumber \\&+ \sum_{j=1}^{N_{ic}}\bigl|H_x(x_j,y_j,0;\theta)\bigr|^2 \nonumber \\&+ \sum_{j=1}^{N_{ic}}\bigl|H_y(x_j,y_j,0;\theta)\bigr|^2]\end{align}

Here, $E_{z_{tar}}(x_j,y_j,0)$ is the prescribed Gaussian profile, and $N_{ic}$ denotes the number of initial condition sampling points distributed over the spatial domain at $t=0$.

\newpage

\section{Implementation}
In the theoretical formulation, expressing Maxwell’s equations as a well-posed initial-boundary value problem reveals an elegant truth: when the equations are formulated correctly, the necessary physical properties emerge naturally. In the considered lossless PEC cavity setting, the PDE, IC and BC trio uniquely determine the solution. This is no coincidence, as Maxwell’s equations automatically ensure energy conservation via Poynting’s theorem, causality via the hyperbolic nature of the wave equations and interface continuity via the electromagnetic field boundary conditions. Therefore, at the \textbf{theoretical stage}, there is no need to define causality or energy loss; such additions would do little more than impose the existing laws again.

However, in the \textbf{implementation stage}, this changes. When training PINNs, the infinite dimensional solution space must be approximated using finite parameters and integral norms must be optimized using sampled estimators. In this discrete, parametric and local optimization based setting, the guarantees provided by theory may be insufficient. Scenarios that contradict physical reality are entirely plausible. For example, the optimization algorithm could violate energy conservation while minimising PDE residuals; the neural network could allow information to flow backwards in time; or field discontinuities could arise at material interfaces. These scenarios do not originate from a deficiency in the theoretical formulation, but from the intrinsic limitations of approximation and sampling based learning in neural networks. Consequently, the main goal at the implementation stage is to translate the physical guarantees implicit in theory into explicit constraints that guide the optimizer.

\subsection{Training Pipeline}
The core of the model is a deep fully connected neural network with 8 hidden layers, each consisting of 128 neurons. During training, to improve gradient flow and to overcome the optimization difficulties of deep networks, a skip connection is applied after every second hidden layer inspired by the ResNet architecture \cite{he2015deepresiduallearningimage}. To ensure robust performance across different geometries, a set of problem-specific engineering approaches has been adopted:

\textbf{Input normalization.} The tanh activation function is symmetric around zero and yields its strongest gradients in this region. As inputs move away from zero, tanh enters saturation and gradients diminish. This imbalance causes some inputs to contribute more to the gradients than others. To prevent this issue, the spatial coordinates are dynamically normalized to the range $[-1, 1]$ depending on the cavity aspect ratio $\beta = b/a$. \cite{LeCun1998} \cite{ioffe2015batchnormalizationacceleratingdeep}

\textbf{Sinusoidal time features.} Electromagnetic wave propagation in a closed cavity inherently exhibits periodic behavior. To explicitly encode this structure, the temporal input is enriched with the sinusoidal features $\sin(2\pi t/T)$ and $\cos(2\pi t/T)$. This approach provides the network with a natural basis for representing oscillatory field evolution. \cite{NEURIPS2020_55053683}

\textbf{Output scaling.} During the non-dimensionalization stage, the relationship between the electric and magnetic field components was defined as $H_0 = E_0/Z_0$, where $Z_0$ is the characteristic impedance of free space. When the neural network's output layer is initially initialized with random weights, all outputs produce values of similar magnitudes. However, physically, the $E_z$ and $H_x, H_y$ components are expected to have different scales. To eliminate this inconsistency, the magnetic field components are scaled by a factor of 0.1 in the output layer. This design facilitates the optimization process by ensuring that the network's starting point lies in a physically meaningful region. \cite{Mandl2023Affine}

\textbf{Two-stage optimization.} The optimization process is based on a two-stage strategy that combines the advantages of first-order and second-order methods. In the first stage, the global structure of the solution is learned for 1500 epochs using the Adam optimizer. In this stage, gradient clipping is used to eliminate the risk of gradient explosion and an LR scheduler that dynamically reduces the learning rate on plateaus where convergence slows down is employed. In the second stage of optimization, starting from the point reached by Adam, high-precision fine-tuning is performed with the L-BFGS optimizer, which uses local curvature information of the loss landscape.

\textbf{Dynamic sampling.} To maximize both the efficiency and the accuracy of the training process, both sampling strategies and loss function weights are managed dynamically. Collocation points, at which the PDE residual is evaluated, are adapted according to the cavity aspect ratio, and their density is increased for narrow geometries.

Similarly, to preserve causality during transitions between time windows, denser sampling is performed in regions close to the beginning of each window. During the early epochs of training, the weight of the loss term representing the initial conditions $({L}_{IC})$ is increased, while in later epochs the weight of the loss term representing the PDE $({L}_{PDE})$ is gradually raised. This integrated approach enables the model to first learn the overall structure of the problem and then refine its solution to achieve full consistency with the governing physical equations.

\subsection{Causality of PINNs}
When the retarded solutions of Maxwell's equations are adopted, the field value at each point depends only on the initial and boundary data within that point’s past light cone. In other words, the field values at $t=1.0$ are formed as a function of the values at $t<1.0$. This behavior is a natural consequence of causality in hyperbolic systems.

Within the PINN framework, however, the neural network treats time as a spatial dimension and attempts to minimize the entire time axis in a single pass. During backpropagation, since the gradient signal from the loss function also incorporates errors at later times, the outputs at early times are modified retroactively. As a result, an error signal observed by the model at a future instant causes an update at a past instant, and causality is violated.

Various approaches have been explored in the literature to prevent this issue. Examples include causal training \cite{wang2022respectingcausalityneedtraining}, which directly encodes causal information flow into the loss design; time marching \cite{Mattey_2022} and causal sweeping \cite{Penwarden_2023}, which arrange training sequentially from past to future; XPINNs-type decompositions \cite{jagtap2020extended}, which split the domain into spacetime subregions; and curriculum learning \cite{10.1145/1553374.1553380}, which aligns optimization by gradually changing difficulty or sampling weights.

In this study, a time-domain decomposition strategy is adopted to prevent causality violation. This strategy is applied to the model through the time marching approach.

\subsubsection{Time Marching}
The time marching method partitions the temporal domain into sequential windows, training each window separately. This approach is inspired by the time stepping concept in FDTD methods, while preserving the mesh-free advantages of PINNs.

The temporal domain $[0, T_{\max}]$ is divided into $N_w$ contiguous windows:
\begin{align}
\mathcal{T} = \{[t_0, t_1], [t_1, t_2], \ldots, [t_{N_w-1}, t_{N_w}]\}
\end{align}
where $t_0 = 0$ and $t_{N_w} = T_{\max}$. For each window $\mathcal{W}_i = [t_i, t_{i+1}]$, a complete training procedure consisting of both Adam and L-BFGS stages is applied independently. In this study, with $T_{\max} = 2.0$ and $\Delta t_{\text{window}} = 0.10$, the temporal domain is partitioned into 20 windows.

The main advantage of this decomposition is that it enables the neural network to handle a more limited temporal complexity within each window. Rather than learning the solution over the entire time span simultaneously, the model focuses on a short interval where the field evolution is more predictable. As the window size decreases, the number of windows increases, and the computational cost grows accordingly. Based on experimental studies, $\Delta t_{\text{window}} = 0.10$ is found to provide a good balance between accuracy and computational cost.

The sampling strategy is adapted for time-windowed training. For each window, fresh collocation points are generated with uniform random distribution over the window's spatio-temporal domain. To improve the solution quality near window boundaries, where the initial condition must be matched precisely, an additional set of collocation points is concentrated near the beginning of each window. Specifically, 25\% extra collocation points are sampled within the first 10\% of each window's temporal span. This denser sampling ensures that the optimizer receives stronger gradient signals for enforcing continuity at window interfaces.

\subsubsection{Sequential Training}
The sequential training strategy ensures that each window is trained conditioned on the completed solution of the previous window. This approach enforces the causality principle and produces physically meaningful solutions.

In the first window $\mathcal{W}_0$, standard PINN training is applied using the analytic Gaussian pulse as the initial condition. For subsequent windows $\mathcal{W}_i$ $(i > 0)$, the trained model from window $\mathcal{W}_{i-1}$ is evaluated at $t = t_i$ to obtain the field values $[E_z, H_x, H_y]$, which then serve as the initial condition for the current window. These predictions are computed in inference mode without gradient tracking and stored as fixed target tensors. During training, the $\mathcal{L}_{\text{IC}}$ loss term penalizes deviations between the current model's predictions at $t = t_i$ and these stored targets, effectively enforcing $C^0$ continuity of the electromagnetic fields across window boundaries.

Furthermore, the network weights from the previously trained window are used to initialize the current window's model parameters. This transfer learning strategy allows each window to inherit the learned representations from the previous window, providing a warm start that improves both training efficiency and solution continuity. Without this initialization, each window would need to learn the field structure from scratch, significantly increasing the total training time.

\subsubsection{Causality-Aware Weighting}
One of the important issues in PINN training is the phenomenon of forward error accumulation, where the error grows as it is propagated forward in time. When time marching is used, even if each window is solved within itself, a small PDE inconsistency at the beginning of the window can spread through derivative terms and become pronounced at mid to late times. For this reason, a causality-aware weighting mechanism is proposed in this study. This mechanism can be considered a simplified and deterministic variant of the causal training approach.

In the causality-aware weighting, for each window, time $t$ is normalized to the $0$ to $1$ range over the interval $[t_{\text{start}}, t_{\text{end}}]$ via $\tau=(t-t_{\text{start}})/(t_{\text{end}}-t_{\text{start}})$, and the PDE residuals are weighted by
\begin{align}
w_c(\tau) = \exp(-\gamma\tau)
\end{align}

In practice, with the choice $\gamma=2$, samples at the beginning of the window $\tau=0$ receive weight $1$, while those at the end $(\tau=1)$ receive approximately $e^{-2}\simeq 0.135$. This creates a weight ratio of approximately $7.4\times$ between the beginning and the end of the window. As a result, stochastic gradient estimates systematically focus on the beginning of the window, optimization suppresses early time PDE inconsistencies first, and an equation-consistent initial regime is established.

The within-window normalization of time is a critical detail and is recommended not to be overlooked. In this way, the numerical meaning of the weighting remains fixed for each window, the gradient scale becomes independent of the window duration, and the causality emphasis is applied consistently across windows.

\subsection{Loss Function Implementation}
In section 1, the theoretical design process of the loss functions was discussed in detail. However, during the implementation stage, it becomes essential to expand the total loss function with additional terms and to make it dynamic. This section discusses how the loss function is transformed during implementation.

\subsubsection{Interface Continuity Loss}
Another component that needs to be added to the total loss function obtained in the theoretical stage during implementation can be defined as the interface continuity loss. Although, in the sequential method, each time window appears as an independent PINN optimization on its own, an additional condition must be enforced to guarantee that the physical solution evolves smoothly over time. This condition, which can be referred to as the interface continuity loss, is necessary to ensure the continuity of the electromagnetic field between successive time windows.

To reliably stitch successive windows, a two-stage continuity mechanism is employed. The first stage is the IC continuity loss applied at the start of each window. In the first window, the initial-condition term enforces the physical initial state by prescribing a Gaussian pulse and zero magnetic field. In subsequent windows, the same term no longer targets an analytic profile, but instead matches the trained PINN solution from the previous window. Accordingly, for successive time windows $\mathcal{W}_i=[t_i,t_{i+1}]$, the $E_z$, $H_x$, and $H_y$ fields at the beginning of window $k$ are penalized in terms of squared error against the fields at the terminal time of window $k-1$. The fields imported from the previous window are \texttt{detach}ed during backpropagation so that they serve as fixed targets; gradients therefore flow only through the parameters of the active window, keeping the training scheme temporally causal.

The second stage is a dedicated interface continuity loss defined on an independent sampling set at the window boundary. This term compares the current window model $u_\theta(x,y,t)=[E_z,H_x,H_y]$ with the previous window predictions $u_{\text{prev}}(x,y,t)$ at interface centered points $(x_{\text{int}}, y_{\text{int}}, t_{\text{int}})$ and the mean squared error is minimized:
\begin{align}L_{\text{int}}(\theta)=\frac{1}{N_{\text{int}}}\sum_{j=1}^{N_{\text{int}}}\big|u_\theta(x_j,y_j,t_i)-u_{\text{prev}}(x_j,y_j,t_i)\big|_2^2\end{align}

This enforces continuity more strongly over a narrow spatiotemporal band around the interface, rather than relying solely on a single IC slice defined at the window start. The weights $w_{\text{ic}}$ and $w_{\text{int}}$ are used to tune the relative influence of these two continuity layers within the overall loss. Since the initial condition in the first window is provided using an analytic reference, the interface loss is not applied there. Both terms are defined on a single time slice on the shared boundary plane $t=t_i$ of successive windows; therefore, they do not include causality-aware weighting.

\subsubsection{Poynting Loss}
At the risk of repetition, it should be stated once more that Maxwell’s equations encode the information of energy conservation, and through their hyperbolic structure, the principle of finite propagation speed; when physically meaningful retarded solutions are adopted, the framework of causality becomes clear. A neural network, however, does not possess this mathematical structure by default. From the network's perspective the solution space is simply an optimization problem, and the laws of physics carry no intrinsic guarantee within it. If the requirement of energy conservation is not imposed as an explicit constraint, the optimizer in practice focuses only on minimizing the $L^2$ loss value. Although time-windowed training slows the tendency to converge toward a trivial solution, it cannot reliably prevent cumulative energy drift. For this reason, during implementation, Poynting’s theorem is incorporated into the loss function as a regularization term $L_{\text{pyt}}$.

A loss implementation based on Poynting theorem can be carried out using two different approaches: global and local. Both approaches rely on the same physical quantities, but they integrate conservation information into optimization in different ways. The global approach produces a single scalar balance from an integral defined over the domain, whereas the local approach enforces the same balance as a residual at each collocation point. Under the non-dimensionalization scheme of Appendix B, the characteristic magnetic field scale $H_0 = E_0/Z_0$ ensures that the electric and magnetic energy contributions are equally weighted. Consequently, the non-dimensional energy density simplifies to:
\begin{align}  
u(x,y,t)  
&= \tfrac{1}{2}\Big(E_z(x,y,t)^2 + H_x(x,y,t)^2 + H_y(x,y,t)^2\Big)  
\end{align}  

and the Poynting vector as
\begin{align}  
\mathbf S(x,y,t)  
&= \mathbf E(x,y,t) \times \mathbf H(x,y,t) \nonumber \\  
&= \big(-E_z(x,y,t)\,H_y(x,y,t),\; E_z(x,y,t)\,H_x(x,y,t)\big)  
\end{align}  

The global approach is based on the integral form of Poynting’s theorem. In an ideal PEC cavity, since the net boundary flux is zero, the total energy is expected to remain constant, and this condition reduces, in non-dimensional form, to
\begin{align}  
\frac{d}{dt} \int_A u(x,y,t) dA = 0  
\end{align}  

By bringing the time derivative inside the integral, the equivalent target
\begin{align}  
\int_A \frac{\partial u(x,y,t)}{\partial t} dA = 0  
\end{align}  

is obtained. The critical point here is the following: the global approach enforces net conservation over the entire domain rather than conservation at every point, meaning that it drives a single scalar balance aggregated over space to zero. In implementation, this spatial integral is discretized as a weighted sum using a chosen numerical integration rule:
\begin{align}  
\int_A \frac{\partial u(x,y,t)}{\partial t} dA  
&\approx  
\sum_{j=1}^{N_q} w_j\frac{\partial u(x_j,y_j,t)}{\partial t}  
\end{align}

In this study, in the global Poynting variant, the spatial integral is evaluated using Gauss-Legendre quadrature, meaning that the nodes $(x_j,y_j)$ and weights $w_j$ are selected according to the Gauss-Legendre rule. As an alternative global discretization, the integral can also be computed on a uniform grid using the trapezoidal rule, in which case the nodes correspond to regular grid points and the weights correspond to the constant and boundary weights produced by the trapezoidal rule. In both cases, the objective is the same: to produce a single scalar net rate of energy change from the expression $\int_A \partial_t u dA$.

In implementation, the term $\partial_t u$ is computed directly via the time derivatives of the fields:

\begin{align} \frac{\partial u(x,y,t)}{\partial t} = & E_z(x,y,t)\,\frac{\partial E_z(x,y,t)}{\partial t} + \nonumber \\ 
&H_x(x,y,t)\,\frac{\partial H_x(x,y,t)}{\partial t} + \nonumber \\ 
&H_y(x,y,t)\,\frac{\partial H_y(x,y,t)}{\partial t} \end{align}

Thus, the global Poynting loss is defined as the squared penalty of this scalar balance:
\begin{align}  
\mathcal{L}_{\text{pyt-global}}  
&=  
\left(  
\sum_{j=1}^{N_q} w_j\frac{\partial u(x_j,y_j,t)}{\partial t}  
\right)^2  
\end{align}  

Because the global formulation controls the total energy behavior at an aggregated level, it is practical from an implementation perspective. However, since it enforces the constraint through a single scalar balance, it is susceptible to the cancellation-of-errors phenomenon. It is possible for $\partial_t u>0$ in one region of the domain while $\partial_t u<0$ occurs in another region, and these violations can partially compensate at the integral level, making the net violation appear small. For this reason, the global approach behaves more like a regularizer that suppresses global drift rather than guaranteeing local consistency. The practical effects of the global variants are discussed quantitatively in the ablation section.

The local approach, on the other hand, directly takes the differential form of Poynting’s theorem as a residual and aims to enforce the conservation law instantaneously at every collocation point within the domain:
\begin{align}  
\frac{\partial u(x,y,t)}{\partial t} + \nabla \cdot \mathbf{S}(x,y,t)  
&= 0  
\end{align}  

This formulation states that the local change in energy density $\big(\partial_t u\big)$ must be balanced by the divergence of the energy flux $\big(\nabla\cdot\mathbf S\big)$ at the same point. In the TMz mode, using $\mathbf S=(-E_zH_y,E_zH_x)$, the divergence term can be written explicitly as
\begin{align} \nabla \cdot \mathbf{S}(x,y,t) = & \frac{\partial}{\partial x}\Big(-E_z(x,y,t)\,H_y(x,y,t)\Big) + \nonumber \\ &\frac{\partial}{\partial y}\Big(E_z(x,y,t)\,H_x(x,y,t)\Big) \end{align}

Accordingly, the local residual that should be physically zero is defined as
\begin{align}  
R_{\text{pyt}}(x,y,t)  
&=  
\frac{\partial u(x,y,t)}{\partial t}  
+  
\nabla \cdot \mathbf{S}(x,y,t)  
\end{align}  

and the loss function is given by the mean squared error of this residual over the collocation set:
\begin{align}  
\mathcal{L}_{\text{pyt-local}}  
&=  
\frac{1}{N_c}\sum_{i=1}^{N_c}  
\left(  
R_{\text{pyt}}(x_i,y_i,t_i\right))^2  
\end{align}  

Since the local approach makes conservation violations visible exactly where they occur, it provides a more pointwise control compared to the global approach. In return, computing $\partial_t$ and spatial derivatives together requires additional derivative evaluations. The impact of the global and local approaches on energy drift and cumulative drift behavior in time-windowed training will not be detailed here to avoid repetition, and will be presented comparatively in the ablation section.

\subsubsection{Final Loss Function and Dynamic Weighting}
The causality-aware mechanisms, interface continuity, and Poynting constraints developed in the previous sections are combined with the core PDE, BC, and IC losses mentioned in Section 1.2 to form the final total loss function.

When the local Poynting approach, which yields the most stable results in this study, is adopted, the total loss $\mathcal{L}_{\text{total}}$ optimized for any time window $\mathcal{W}_i$ is expressed as a dynamically weighted sum of all these components:
$$\mathcal{L}_{\text{total}} = w_{\text{pde}}\mathcal{L}_{\text{PDE}} + w_{\text{bc}}\mathcal{L}_{\text{BC}} + w_{\text{ic}}\mathcal{L}_{\text{IC}} + w_{\text{int}}\mathcal{L}_{\text{INT}} + w_{\text{pyt}}\mathcal{L}_{\text{pyt-local}}$$

It is important to note that the term $\mathcal{L}_{\text{PDE}}$ is already modified internally by the causality-aware weighting $w_c(\tau)=\exp(-\gamma\tau)$ described in Section 2.2.3.

This equation does not represent a static sum. The weight vector $\mathbf{w} = [w_{\text{pde}}, w_{\text{bc}}, w_{\text{ic}}, \dots]$ follows a dynamic schedule depending on the two-stage training strategy (Adam and L-BFGS) summarized in Section 2.1. For example, in the early epochs of Adam optimization, the weight $w_{\text{ic}}$ is increased so that the model focuses first on the initial conditions, while regularizer terms such as $w_{\text{pyt}}$ are initialized with a lower weight. As optimization progresses, these weights are updated to enforce tighter compliance with the PDE and physical conservation laws. This dynamic loss function guides the loss landscape toward stable convergence and enables the model to obtain physically consistent solutions.

\newpage

\section{Results and Ablation}
\subsection{Error Metrics and Validation}
In this study, a high-resolution Finite-Difference Time-Domain (FDTD) simulation is used as a reference solution to evaluate the accuracy of the field solutions produced by the trained PINN model. The FDTD solver is not included in any stage of the training process; it is run only after training is completed, in order to generate reference data under the same physical problem definition and to perform a quantitative comparison with the PINN outputs.

The reference FDTD solution is generated on a Yee staggered grid. The time step is chosen as $dt=0.5\cdot\min(dx,dy)$ which satisfies the CFL stability condition for the 2D wave equation, and the total number of steps is set to $steps=\frac{t_{\max}}{dt}+1$. In the analyses, 41 equally spaced snapshot times are used over the interval $t\in[0,2.0]$. 

To maintain readability, tables contain a limited number of time steps, but the \textbf{Avg.} rows indicating Average values are calculated by taking the average of the metrics at 41 time steps.

The FDTD solver advances and computes all field components $E_z$, $H_x$ and $H_y$ over time in a manner consistent with the staggered-grid structure of the problem. For energy-based validation and a fair comparison with the PINN, fields are temporally synchronized using a leapfrog-consistent half-step averaging so that E and H are evaluated at the same effective time for energy calculations. 

FDTD provides the $(x,y,t)$-dependent reference field data $U_{\mathrm{ref}}= E_{z{ref}},H_{x{ref}},H_{y{ref}}$ at the snapshot times, along with the corresponding energy evolution, forming the baseline reference for the error metrics defined in the remainder of this section.

Two complementary field accuracy metrics are reported. The primary metric is Normalized Root Mean Squared Error. NRMSE is selected for its robustness in wave problems where relative metrics can exhibit artificial spikes due to small denominators in nodal regions. As a secondary indicator, the relative $L_2$ error on the combined vector field is also reported to capture global, scale-aware deviations. Formal definitions, including the normalization convention and the aggregated multi-component form, are provided in Appendix C.

In addition to pointwise field fidelity, energy behavior is evaluated to verify long-time physical consistency. The analysis includes (i) instantaneous energy mismatch between PINN and FDTD, (ii) each method’s internal energy conservation relative to its own initial energy, and (iii) time-window interface continuity in terms of energy jumps between consecutive windows in the sequential training pipeline. These metrics are computed on a uniform collocated evaluation grid to enable fair side-by-side post-processing of PINN and FDTD samples.

Finally, to ensure that conclusions are not an artifact of the collocated evaluation grid, an additional native-grid validation is performed based on the Yee-grid discrete energy functional. In this analysis, the PINN solution is transferred onto the same Yee grid locations used by FDTD, and the discrete energy trajectories are compared directly. Summary statistics of the resulting discrete energy mismatch are reported alongside the continuous-grid energy metrics. Full definitions are given in Appendix C.

\subsection{Results}

Multiple versions of the PINN model developed in this study were trained under three scenarios and with different implementation approaches. The main version from which the other versions are derived will be referred to as \textbf{pinnA}.

One of the most fundamental challenges faced by time marching PINN models in the literature is the error accumulation problem. \cite{CHEN2024111423} In the continuous lossless PEC setting, total energy is constant; therefore deviations in discrete solvers like FDTD are treated as numerical error. 

In neural network applications for electromagnetic problems, this error is often characterized as energy drift and most commonly manifests as decay. The underlying reason is the tendency of neural networks to converge toward trivial solutions. Rather than amplifying field magnitudes, the network tends to suppress them; this leads to amplitude damping and energy loss. \cite{CHANG2024112791}

Although the time marching strategy is successful in preventing causality collapse, it causes energy drift to become cumulative. This situation can be seen as a natural consequence of how the method operates. In the time marching method, the terminal values of each time window are treated as the initial condition for the next window. This causes every small error formed in the previous stage to be inherited by the subsequent window. Therefore, energy drift becomes cumulative and deepens as time progresses.

The most distinctive property of \textbf{pinnA} is that it does not exhibit a cumulative energy drift pattern. Inspection of the results indicates that when the model experiences an energy drop within a given window, it compensates for that drop in subsequent windows and brings the solution back toward the reference energy level. This non-monotonic error behavior suggests that the model is not only minimizing error, but also leveraging physical conservation laws to correct itself.

\subsubsection{Main Model}
\begin{figure*}[ht]\centering 
	\includegraphics[width=\linewidth]{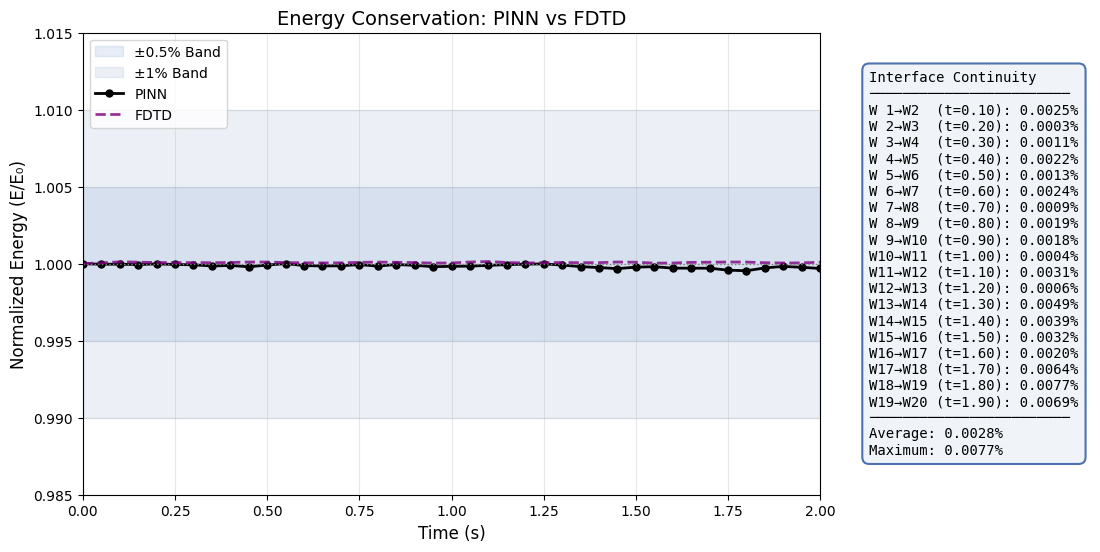}
	\caption{Energy Conservation: PINN vs FDTD}
	\label{fig:18_2_333}
\end{figure*}

\textbf{pinnA} is a fully-connected neural network trained along the time axis using a time marching approach. The network architecture consists of 8 hidden layers with 128 neurons per layer. The computational domain is defined by $x \in [0, 1]$ and $y \in [0, \beta]$, where $\beta = 1.0$ represents the aspect ratio of the cavity. The global time interval $t \in [0, 2.0]$ is partitioned into 20 consecutive windows, each with width $\Delta t_{\text{window}} = 0.10$, and trained independently. The initial condition is specified by a Gaussian pulse with amplitude $A = 1.0$ and width $\sigma = 0.1$. For each window, the training data consists of $100000/\beta$ collocation points, $20000/\beta$ boundary points, and $3000/\beta$ initial condition points. The loss function is defined as a weighted sum of five terms: PDE residual ($w_{\text{PDE}} = 1.0$), boundary conditions ($w_{\text{BC}} = 50.0/\beta$), initial condition ($w_{\text{IC}} = 100.0$), window-interface continuity ($w_{\text{interface}} = 75.0$), and Poynting-vector-based energy conservation ($w_{\text{Poynting}} = 10.0$). Each window is trained using 1500 epochs of the Adam optimizer with learning rate $10^{-3}$, followed by 300 epochs of L-BFGS with learning rate $0.5$.

The FDTD solution is not incorporated into the training process in any form; therefore, the risk of overfitting to FDTD data is eliminated by construction. During the analysis stage, an FDTD solver is used purely as a reference to quantitatively evaluate the accuracy of \textbf{pinnA}.

Field-level accuracy of \textbf{pinnA} relative to FDTD is assessed using the Normalized Root Mean Square Error (NRMSE) and the relative $L^2$ norms defined in the previous section. The results for selected time steps are reported in the Table~\ref{tab:accuracy_metrics}. The reported average values are computed not only over the time steps shown in the Table~\ref{tab:accuracy_metrics}, but as an average over all time steps used in the analysis, $t = [0.00, 0.05, 0.10, 0.15, \ldots, 2.00]$.

As shown in the Table~\ref{tab:accuracy_metrics}, the NRMSE values for all three field components remain on the order of 0.1–0.2\% throughout the full time interval. Even the relative $L^2$ error reaches only about $1.33\%$ in the worst case and stays around $1.01\%$ on average. 

\begin{table*}[t]
\centering
\caption{pinnA: Accuracy metrics over time.}
\label{tab:accuracy_metrics}
\begin{tabular}{l
                S[table-format=1.2]
                S[table-format=1.2]
                S[table-format=1.2]
                S[table-format=1.2]
                S[table-format=1.2]}
\toprule
Time & {NRMSE\_Ez (\%)} & {NRMSE\_Hx (\%)} & {NRMSE\_Hy (\%)} & {NRMSE\_Total (\%)} & {$L_2$ Total (\%)} \\
\midrule
$t=0.0$ & 0.01 & 0.02 & 0.01 & 0.01 & 0.08 \\
$t=0.5$ & 0.17 & 0.16 & 0.12 & 0.10 & 1.05 \\
$t=1.0$ & 0.12 & 0.17 & 0.15 & 0.08 & 0.85 \\
$t=1.5$ & 0.17 & 0.20 & 0.18 & 0.10 & 1.14 \\
$t=2.0$ & 0.19 & 0.25 & 0.26 & 0.12 & 1.33 \\
\midrule
Avg.   & 0.16 & 0.17 & 0.15 & 0.09 & 1.01 \\
\bottomrule
\end{tabular}
\end{table*}

\begin{table*}[t]
\centering
\caption{pinnA: Energy metrics over time.}
\label{tab:energy_metrics}
\begin{tabular}{l
                S[table-format=1.6]
                S[table-format=1.6]
                S[scientific-notation=true, table-format=1.3e-1]
                S[table-format=1.3]
                S[table-format=+1.3]
                S[table-format=1.3]}
\toprule
Time & {PINN Energy} & {FDTD Energy} & {Abs. Error} & {Rel. Error (\%)} & {Cons. PINN (\%)} & {Cons. FDTD (\%)} \\
\midrule
$t=0.0$ & 0.015708 & 0.015708 & 1.482e-07 & 0.001 &  0.000 & 0.000 \\
$t=0.2$ & 0.015707 & 0.015709 & 2.047e-06 & 0.013 & -0.005 & 0.007 \\
$t=0.5$ & 0.015706 & 0.015710 & 3.210e-06 & 0.020 & -0.009 & 0.011 \\
$t=0.8$ & 0.015707 & 0.015709 & 2.450e-06 & 0.016 & -0.006 & 0.009 \\
$t=1.0$ & 0.015705 & 0.015709 & 3.535e-06 & 0.023 & -0.017 & 0.005 \\
$t=1.2$ & 0.015708 & 0.015709 & 1.346e-06 & 0.009 & -0.001 & 0.007 \\
$t=1.5$ & 0.015704 & 0.015709 & 5.139e-06 & 0.033 & -0.022 & 0.010 \\
$t=1.8$ & 0.015701 & 0.015710 & 8.406e-06 & 0.054 & -0.042 & 0.011 \\
$t=2.0$ & 0.015703 & 0.015709 & 6.503e-06 & 0.041 & -0.031 & 0.010 \\
\midrule
Avg.   & 0.015705 & 0.015709 & 3.759e-06 & 0.024 & 0.015 & 0.008 \\
\bottomrule
\end{tabular}
\end{table*}

While accuracy metrics capture the convergence capability of the neural network, they are not sufficient on their own. To verify that the solution is consistent with physical behavior, the energy metrics defined in the previous part are also employed.

Since the values in the Table~\ref{tab:energy_metrics} and the figure are computed on a collocated grid for the PINN solution, FDTD and PINN are additionally evaluated on a common Yee grid to confirm consistency. Under the staggered-grid layout, the mean absolute difference over time is $3.74 \times 10^{-6}$, and the maximum difference is $9.07 \times 10^{-6}$. When normalized by the reference energy, the mean relative mismatch is only $0.024\%$, and the maximum mismatch is $0.057\%$. In other words, the discrete energy functional of the PINN tracks the discrete FDTD energy across the entire spatio-temporal domain at a level well below $10^{-3}$ relative error.
\begin{figure}[ht]\centering
	\includegraphics[width=\linewidth]{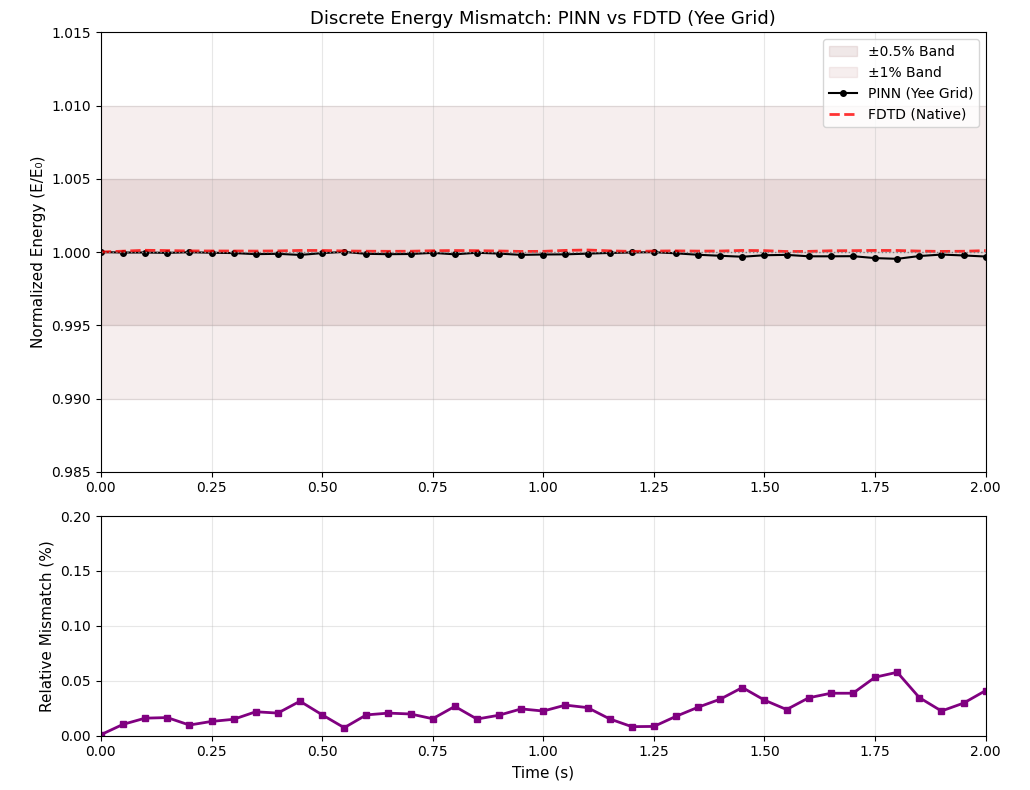}
	\caption{Yee Grid Energy Mismatch}
	\label{fig:18_2_33}
\end{figure}

{
\sisetup{
  scientific-notation=true,
  output-exponent-marker=e,
  round-mode=none
}
\begin{table}[t]
\centering
\caption{Discrete Energy Mismatch}
\small
\begin{tabular*}{\columnwidth}{@{\extracolsep{\fill}}l r@{}}
\toprule
\textbf{Metric} & \textbf{Value} \\
\midrule
Mean Absolute Difference & \num{3.743911e-06} \\
Max Absolute Difference  & \num{9.074444e-06} \\
Mean Relative Difference (\%) & 0.0238 \\
Max Relative Difference (\%)  & 0.0578 \\
PINN Energy Variation (\%)    & 0.0472 \\
FDTD Energy Variation (\%)    & 0.0139 \\
\bottomrule
\end{tabular*}
\end{table}
}

\subsubsection{Lossy Scenario}
To test whether the stable field and energy behavior exhibited by \textbf{pinnA} in the lossless PEC cavity is preserved under lossy medium conditions, a lossy variant was trained using the same time-marching architecture and the same configuration.

The differences between \textbf{pinnA} and the lossy variant are not limited to the medium model; the lossy physics directly affects both the PDE residual and the Poynting constraint formulations. The conductivity parameter is set to $\eta = \sigma_c/\varepsilon_0 = 0.5$ in non-dimensional form. The conductivity term $-\eta E_z$ is incorporated into the PDE residual derived from Ampere's law:
$$R_1 = \frac{\partial H_y}{\partial x} - \frac{\partial H_x}{\partial y} - \eta E_z - \frac{\partial E_z}{\partial t}$$

Similarly, the Poynting theorem residual is extended to include the Joule heating loss:
$$R_{\text{pyt}} = \frac{\partial u}{\partial t} + \nabla \cdot \mathbf{S} + \eta E_z^2$$

In this formulation, the term $\eta E_z^2$ represents the rate at which electromagnetic energy is converted into heat in the conductive medium.

A difference also exists in the loss function configuration. While the Poynting loss weight is $w_{\text{pyt}} = 10$ in \textbf{pinnA}, this value is set to $w_{\text{pyt}} = 15$ in the lossy variant. 

\begin{table}[htbp]
\centering
\caption{Lossy Scenario Results}
\small
\begin{tabular}{@{}l ccccc@{}}
\toprule
\textbf{Time} & \textbf{NRMSE} & \textbf{L2} & \textbf{FDTD.E} & \textbf{pinn.E} & \textbf{Rel.Enrgy.Err} \\
\midrule
t=0.0 & 0.01\% & 0.08\% & 0.015708 & 0.015708 & 0.002\% \\
t=0.5 & 0.09\% & 1.04\% & 0.012601 & 0.012582 & 0.154\% \\
t=1.0 & 0.06\% & 0.91\% & 0.009353 & 0.009336 & 0.187\% \\
t=1.5 & 0.08\% & 1.26\% & 0.007390 & 0.007374 & 0.220\% \\
t=2.0 & 0.10\% & 1.80\% & 0.005858 & 0.005847 & 0.199\% \\
\midrule
\textbf{Avg.} & \textbf{0.08\%} & \textbf{1.11\%} & \textbf{0.009958} & \textbf{0.009941} & \textbf{0.176\%} \\
\bottomrule
\end{tabular}
\end{table}
In terms of field accuracy, the lossy variant performs at the same order of magnitude as \textbf{pinnA}. The average $\text{NRMSE}_{\text{Total}}$ is measured as 0.08\% in both scenarios, and the average NRMSE values for the $E_z$, $H_x$, and $H_y$ components remain on the order of one per mille throughout the entire time interval. The relative $L^2$ error averages 1.11\% in the lossy scenario, remaining close to the lossless result. These results indicate that, despite the addition of the conductivity term and the reduction in the Poynting weight, the network retains its capacity to produce a consistent field solution across the spatio-temporal domain.

The energy metrics provide a more critical physical validation. The initial energy nearly coincides with FDTD in both scenarios, with the relative initial mismatch not exceeding 0.002\%. As time progresses, the PINN and FDTD energies drift together in a consistent manner. The success criterion here is not that the global energy remains constant, but that the expected dissipation curve under the defined lossy dynamics is accurately tracked. The energy decreases from 0.0157 to 0.0059, corresponding to an approximately 63\% reduction by $t = 2.0$, and the PINN tracks this decay profile with a relative error below 0.2\%.

Window transitions also remain controlled in the lossy scenario. While \textbf{pinnA} exhibits an average energy jump of 0.0028\% and a maximum jump of 0.0077\%, the corresponding values in the lossy variant are 0.0027\% and 0.0095\%, respectively. These results support the conclusion that the interface continuity mechanism preserves inter-window coherence in the lossy regime as well, keeping time-marching-induced artificial energy jumps at negligible levels.

\subsubsection{Four-Window Variant}
To examine the impact of time marching on \textbf{pinnA}, the training was repeated with $window_size = 0.50$, resulting in a four-window setup instead of 20 windows. Although reducing the number of windows increases the error level, the absolute performance of \textbf{pinnA} remains in a strong range. 

\begin{table}[htbp]
\centering
\caption{Four-Window Variant of pinnA}
\small
\begin{tabular}{@{}l ccccc@{}}
\toprule
\textbf{Time} & \textbf{NRMSE} & \textbf{L2} & \textbf{FDTD.E} & \textbf{pinn.E} & \textbf{Rel.Enrgy.Err} \\
\midrule
t=0.0 & 0.02\% & 0.27\% & 0.015708 & 0.015703 & 0.028\% \\
t=0.5 & 0.47\% & 5.11\% & 0.015782 & 0.015662 & 0.765\% \\
t=1.0 & 0.47\% & 5.24\% & 0.015603 & 0.015732 & 0.824\% \\
t=1.5 & 0.52\% & 5.79\% & 0.015777 & 0.015715 & 0.395\% \\
t=2.0 & 0.59\% & 6.59\% & 0.015658 & 0.015672 & 0.095\% \\
\midrule
\textbf{Avg.} & \textbf{0.42\%} & \textbf{4.60\%} & \textbf{0.015705} & \textbf{0.015697} & \textbf{0.421\%} \\
\bottomrule
\end{tabular}
\end{table}

The average $NRMSE_{\text{Total}}$ is measured as $0.42\%$ and the average $L2_{\text{Total}}$ as $4.60\%$. On the energy side, agreement with FDTD is maintained, with the mean relative energy error remaining at $0.421\%$. Despite a relatively large initial energy mismatch of $0.028\%$ compared to \textbf{pinnA}, the relative energy error decreases to $0.095\%$ at $t=2.0$. In this configuration, even with fewer windows, the energy jumps remain at $0.0187\%$ on average and $0.0322\%$ at worst. This indicates that the local Poynting constraint and the interface-based regularization, which are analyzed in detail in the ablation section, keep the relative energy error below 0.5\% even when wider time windows are used.

\subsubsection{High Frequency Scenario}
To test the robustness of pinnA against spectral bias, tests were also conducted in the high frequency region. In the pinnA configuration, the Gaussian pulse width $\sigma$ = 0.1 was set, which corresponds to a wave structure of approximately 7.8 periods in the medium frequency regime. In the high frequency variant pinnHF, $\sigma$ = 0.05 was set, thereby increasing the maximum dominant frequency from 3.902 to 8.780 and the number of periods to approximately 17.6.

\begin{table}[h]
\centering
\caption{Comparison of pinnA and pinnHF configurations.}
\begin{tabular}{lcc}
\hline
Metric & pinnA & pinnHF \\
\hline
NRMSE (avg) & 0.09\% & 0.10\% \\
$L^2$ Error (avg) & 1.01\% & 2.21\% \\
Energy Rel. Error & 0.02\% & 0.21\% \\
Energy Mismatch (Yee Grid) & 0.02\% & 0.21\% \\
Max. Dominant Freq. & 3.902 & 8.780 \\
Max. Number of Periods & $\sim$7.8 & $\sim$17.6 \\
\hline
\end{tabular}
\end{table}

When the results are examined, pinnHF demonstrated strong performance despite the frequency content increasing by more than twofold. The average NRMSE was measured as 0.10\%, the average $L^2$ error as 2.21\%, and the relative energy error as 0.21\%. These results indicate that the proposed methodology works effectively in the high frequency regime as well, despite the spectral bias phenomenon where neural networks tend to learn low frequency components faster. The fact that error metrics remain within acceptable limits despite the significant increase in frequency content suggests that the time marching strategy and Poynting based regularization can partially compensate for spectral bias.

When the frequency band is increased further, it was observed that the model's success diminishes. For $\sigma$ = 0.03, when the maximum number of periods is approximately 29.9, the average relative energy error reaches 6.862\%. A large portion of the relative error occurs at the initial instant when the Poynting regularizer operates with low weight. With the active engagement of the Poynting regularizer, a decrease in energy drift was observed.

It can be stated that $\sigma$ = 0.05 represents a performance boundary for pinnA, beyond which the current model loses its stability. However, it has been understood that the spectral bias problem can be substantially mitigated through improvements to the Poynting regularizer. A spectral bias resistant PINN configuration is targeted in future work.

\subsubsection{The Parenthesis Effect}
In this study, an additional control variant was observed that produced a noteworthy divergence in training dynamics. This variant will hereafter be referred to as \textbf{pinnP}.

\textbf{pinnP} was obtained by copying the \textbf{pinnA} folder exactly and retraining with the same model. The only difference is a single line in the Poynting loss function. Apart from this, all files are identical character by character.

This difference can be summarized by the following two lines. In the \textbf{pinnA}, the divergence term is written with parentheses grouping two subexpressions:

\begin{lstlisting}
div_S = (-Ez_x * Hy - Ez * Hy_x) + (Ez_y * Hx + Ez * Hx_y)
\end{lstlisting}

In the \textbf{pinnP}, the same line is written without parentheses:

\begin{lstlisting}
div_S = -Ez_x * Hy - Ez * Hy_x + Ez_y * Hx + Ez * Hx_y
\end{lstlisting}
Algebraically, the two expressions are equivalent. Nevertheless, post-training metrics exhibited a noticeable separation, particularly in energy behavior. This observation is referred to as \textbf{the parenthesis effect}.

Under the 20-window setup, both \textbf{pinnA} and \textbf{pinnP} avoid pronounced cumulative energy drift. However, \textbf{pinnP} shows weaker energy conservation; the normalized energy curve oscillated over a wider band and relative mismatch reached higher values.

The divergence becomes more apparent in the 4-window setup. Here, \textbf{pinnA} experienced a transient overshoot around $t \approx 0.5$, yet recovered toward the reference curve later. Its final relative energy error at $t = 2.0$ is approximately 0.236\%. 

In contrast, \textbf{pinnP} did not exhibit such overshoot; instead, it produced a more pronounced drift pattern, with relative energy error remaining at approximately 0.416\% att $t = 2.0$.

Figure 4 compares the normalized energy curves and relative mismatch evolution for the four-window setup.
\begin{figure}[htbp]\centering
	\includegraphics[width=\linewidth]{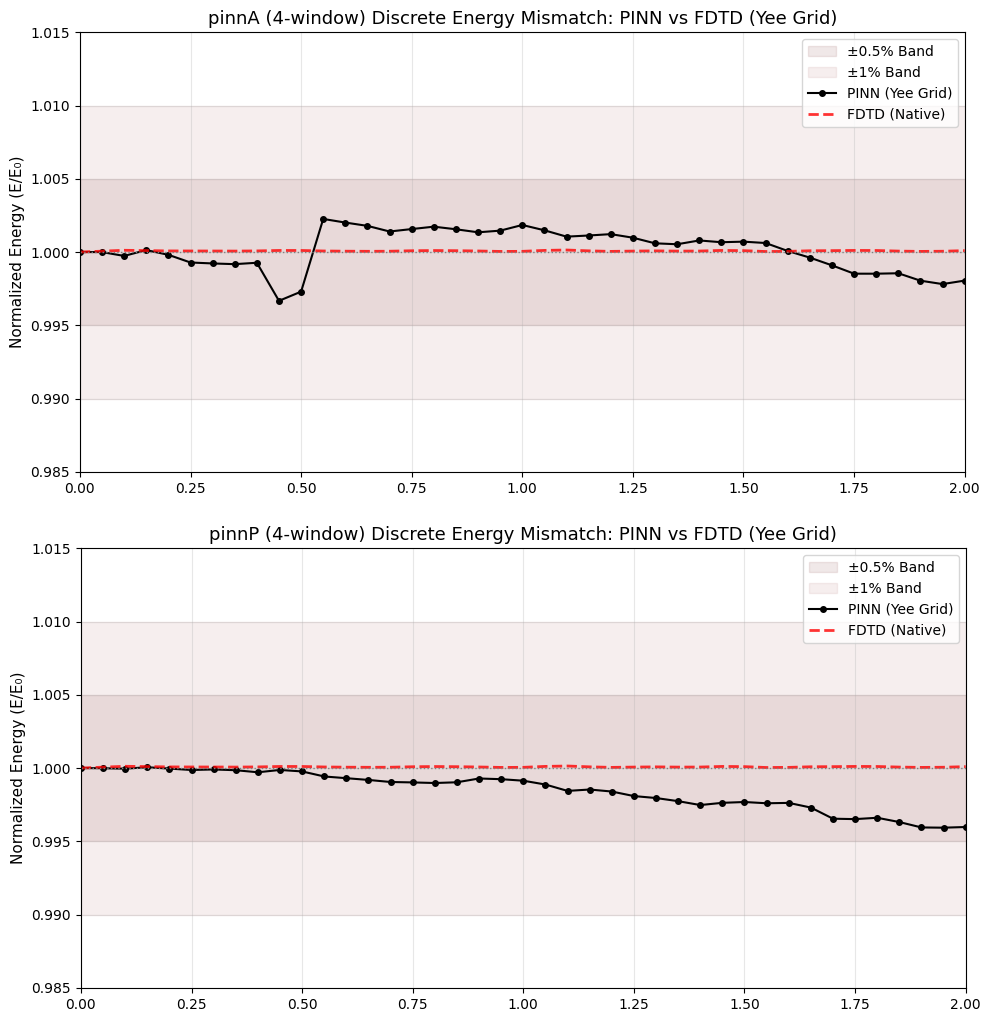}
	\caption{pinnA vs pinnP (four-window versions)}
	\label{fig:18_35}
\end{figure}

A plausible contributing factor is floating-point non-associativity and its interaction with the autodiff graph. However, the point here is not the technical cause but the practical consequences for PINN models.

In classical methods such as FDTD or FEM, writing algebraically equivalent expressions differently produces negligible differences at rounding-error level. PINNs, however, are built on gradient-based optimization, where every loss term influences parameters through the computational graph. The graph topology can vary with how expressions are written, so mathematically identical formulations may exhibit different optimization dynamics.

This observation highlights an important consideration: code-level modifications that appear algebraically inconsequential can affect long-term model behavior unexpectedly. The parenthesis effect should therefore be interpreted not as an ablation experiment, but as an implementation note on the distinctive sensitivity of PINN-based solvers compared to classical methods.

\subsection{Ablation}
\begin{table*}[t]
\centering
\caption{Ablation Study: Energy and Accuracy Metrics Comparison}
\label{tab:ablation}
\small
\begin{tabular*}{\textwidth}{@{\extracolsep{\fill}}ll cccccc cc@{}}
\toprule
& & \multicolumn{6}{c}{\textbf{Energy Metrics}} & \multicolumn{2}{c}{\textbf{Accuracy Metrics}} \\
\cmidrule(lr){3-8} \cmidrule(lr){9-10}
\textbf{Model} & \textbf{Time} & PINN.E & FDTD.E & Abs.Err & Rel.Err & Cons.PINN & Cons.FDTD & NRMSE$_{Tot}$ & L2$_{Tot}$ \\
\midrule
\multirow{6}{*}{pinnA}
 & t=0.0 & 0.015708 & 0.015708 & 1.482e-07 & 0.001\% & 0.000\%  & 0.000\%  & 0.01\% & 0.08\% \\
 & t=0.5 & 0.015706 & 0.015710 & 3.210e-06 & 0.020\% & -0.009\% & 0.011\%  & 0.10\% & 1.05\% \\
 & t=1.0 & 0.015705 & 0.015709 & 3.535e-06 & 0.023\% & -0.017\% & 0.005\%  & 0.08\% & 0.85\% \\
 & t=1.5 & 0.015704 & 0.015709 & 5.139e-06 & 0.033\% & -0.022\% & 0.010\%  & 0.10\% & 1.14\% \\
 & t=2.0 & 0.015703 & 0.015709 & 6.503e-06 & 0.041\% & -0.031\% & 0.010\%  & 0.12\% & 1.33\% \\
\cmidrule(l){2-10}
 & \textbf{Avg.} & \textbf{0.015705} & \textbf{0.015709} & \textbf{3.759e-06} & \textbf{0.024\%} & \textbf{0.015\%} & \textbf{0.008\%} & \textbf{0.09\%} & \textbf{1.01\%} \\
\midrule
\multirow{6}{*}{pinnGP}
 & t=0.0 & 0.015707 & 0.015708 & 6.120e-07 & 0.004\% & 0.000\%  & 0.000\%  & 0.01\% & 0.07\% \\
 & t=0.5 & 0.015701 & 0.015710 & 9.124e-06 & 0.058\% & -0.043\% & 0.011\%  & 0.10\% & 1.07\% \\
 & t=1.0 & 0.015694 & 0.015709 & 1.493e-05 & 0.095\% & -0.086\% & 0.005\%  & 0.08\% & 0.90\% \\
 & t=1.5 & 0.015692 & 0.015709 & 1.791e-05 & 0.114\% & -0.100\% & 0.010\%  & 0.11\% & 1.19\% \\
 & t=2.0 & 0.015685 & 0.015709 & 2.445e-05 & 0.156\% & -0.142\% & 0.010\%  & 0.14\% & 1.51\% \\
\cmidrule(l){2-10}
 & \textbf{Avg.} & \textbf{0.015695} & \textbf{0.015709} & \textbf{1.385e-05} & \textbf{0.088\%} & \textbf{0.077\%} & \textbf{0.008\%} & \textbf{0.10\%} & \textbf{1.06\%} \\
\midrule
\multirow{6}{*}{pinnWP}
 & t=0.0 & 0.015708 & 0.015708 & 2.020e-07 & 0.001\% & 0.000\%  & 0.000\%  & 0.01\% & 0.07\% \\
 & t=0.5 & 0.015706 & 0.015710 & 3.184e-06 & 0.020\% & -0.011\% & 0.011\%  & 0.10\% & 1.05\% \\
 & t=1.0 & 0.015701 & 0.015709 & 7.296e-06 & 0.046\% & -0.043\% & 0.005\%  & 0.08\% & 0.86\% \\
 & t=1.5 & 0.015699 & 0.015709 & 1.067e-05 & 0.068\% & -0.060\% & 0.010\%  & 0.10\% & 1.14\% \\
 & t=2.0 & 0.015693 & 0.015709 & 1.627e-05 & 0.104\% & -0.095\% & 0.010\%  & 0.11\% & 1.24\% \\
\cmidrule(l){2-10}
 & \textbf{Avg.} & \textbf{0.015702} & \textbf{0.015709} & \textbf{6.933e-06} & \textbf{0.044\%} & \textbf{0.039\%} & \textbf{0.008\%} & \textbf{0.09\%} & \textbf{0.99\%} \\
\bottomrule
\end{tabular*}
\end{table*}

\begin{figure*}[ht]\centering 
	\includegraphics[width=\linewidth]{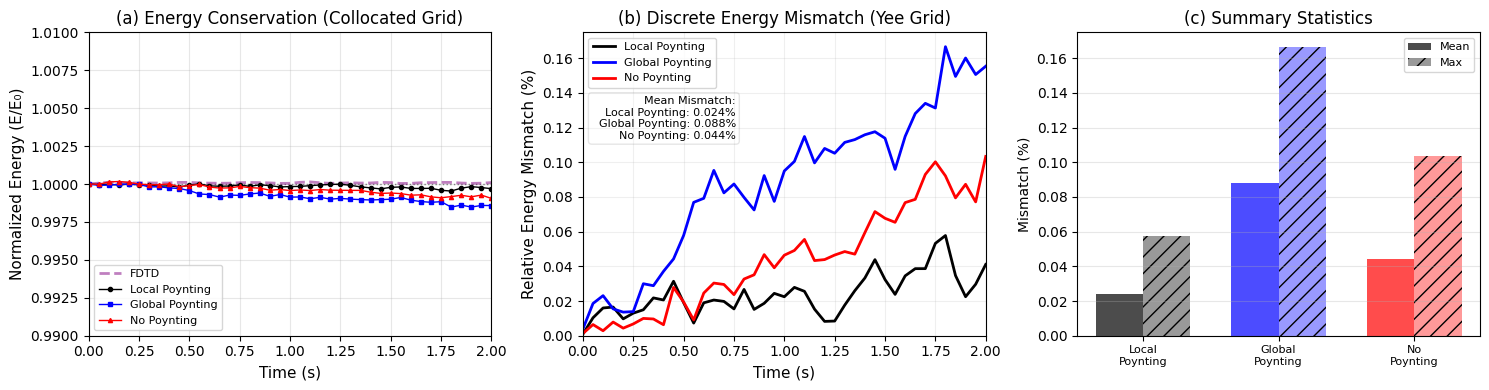}
	\caption{Ablation: pinnA (local Poynting), pinnGP (global Poynting), and pinnWP (no Poynting constraint) compared against FDTD.}
	\label{fig:555}
\end{figure*}

The purpose of this ablation study is to isolate the cumulative energy drift behavior observed in the time-windowed training approach while keeping all other components fixed. To this end, three variants were trained using the same PINN architecture, time-marching scheme, sampling strategy, and loss weights. The variants differ exclusively in how the Poynting-based energy constraint $\mathcal{L}_{\text{pyt}}$ is formulated; all other loss components (PDE, BC, IC, interface continuity) remain identical. Since all aspects of training other than this single design choice are held constant, the differences reported in Table~\ref{tab:ablation} can be attributed directly to the energy-regularization strategy.

\paragraph{pinnA: Local Poynting Residual.} The differential form of Poynting's theorem is enforced as an additional conservation equation at the same collocation points as the Maxwell residuals. The condition $\partial_t u + \nabla \cdot \mathbf{S} = 0$ is penalized at every point in the spatio-temporal domain. Since the relationship between energy flux and energy density is locally constrained, each time window can establish its own internal energy balance, and the drift that would otherwise accumulate across window boundaries is largely suppressed.

\paragraph{pinnGP: Global Poynting Regularizer.} Instead of pointwise enforcement, \textbf{pinnGP} constrains the rate of change of total energy integrated over the domain, approximated via Gauss--Legendre quadrature with $N_q$ nodes and weights $\{(x_j, y_j), w_j\}_{j=1}^{N_q}$:
\begin{align}
\mathcal{L}_{\text{pyt-global}} = \left( \sum_{j=1}^{N_q} w_j \frac{\partial u}{\partial t}(x_j, y_j, t) \right)^2
\end{align}

This formulation produces a single scalar constraint per time instant rather than $N_c$ independent constraints distributed across the domain. Consequently, excess energy in one region can compensate for a deficit elsewhere at the integral level, masking local violations through cancellation of errors. As discussed below, this mechanism does not merely limit the effectiveness of the constraint; it can actively degrade energy behavior relative to having no constraint at all.

\paragraph{pinnWP: No Poynting Constraint.} The energy-conservation term is omitted entirely ($\mathcal{L}_{\text{pyt}} = 0$). This variant serves as a control to isolate the effect of energy regularization from other components of the training pipeline.

\medskip

The accuracy metrics in Table~\ref{tab:ablation} indicate that all three variants achieve comparable field accuracy against the FDTD reference. Both NRMSE and $L^2$ errors remain at the sub-percent level along the time axis, with only minor fluctuations among the three models. This similarity is important: the differences in energy behavior cannot be attributed to one model producing better or worse field solutions, but are a direct consequence of the energy-regularization design.

When the energy metrics are examined, a clear hierarchy emerges. \textbf{pinnA} exhibits the most consistent energy behavior, keeping the normalized total energy within a narrow band and achieving the lowest conservation error both on average and at extreme values. \textbf{pinnWP}, despite having no energy constraint, occupies an intermediate position: cumulative drift is visible as the normalized energy decreases monotonically, but the magnitude remains moderate. The counterintuitive result is that \textbf{pinnGP}, the only variant with a global energy constraint, performs worst in every energy metric. Its average relative error (0.088\%) is roughly double that of \textbf{pinnWP} (0.044\%) and nearly four times that of \textbf{pinnA} (0.024\%). The Cons.PINN column makes the contrast even more explicit: by $t = 2.0$, \textbf{pinnGP} has drifted to $-0.142\%$, compared with $-0.095\%$ for \textbf{pinnWP} and only $-0.031\%$ for \textbf{pinnA}.

This outcome can be understood through the cancellation-of-errors mechanism inherent to the global formulation. Because the global loss penalizes only the net integral $\sum w_j \partial_t u$, the optimizer can satisfy the constraint by allowing positive and negative $\partial_t u$ violations to cancel across the domain. The resulting locally inconsistent energy distributions do not trigger a penalty at the loss level, yet they distort the solution in ways that compound across window boundaries. In effect, the global constraint introduces an additional optimization pathway that permits, and may even encourage, physically inconsistent local energy distributions, producing worse cumulative drift than the unconstrained baseline.

To verify that this finding is not an artifact of the collocated evaluation grid, the analysis is repeated on the native Yee grid discretization of the FDTD solver. The PINN fields are sampled at Yee grid locations and discrete energy time series are computed using the same discrete energy functional as FDTD. The resulting ranking is fully consistent: \textbf{pinnA} produces the lowest mean and maximum differences as well as the lowest energy variation; these discrepancies increase for \textbf{pinnWP} and reach their highest level for \textbf{pinnGP}.

\begin{table}[t]
\centering
\caption{Yee Grid Discrete Energy Analysis}
\small
\begin{tabular}{@{}l ccc@{}}
\toprule
\textbf{Metric} & \textbf{pinnA} & \textbf{pinnGP} & \textbf{pinnWP} \\
\midrule
Mean Abs. Dif.      & 3.743911e-06 & 1.383613e-05 & 6.920973e-06 \\
Max Abs. Dif.       & 9.074444e-06 & 2.617539e-05 & 1.623682e-05 \\
Mean Rel. Dif. (\%) & 0.0238\%     & 0.0881\%     & 0.0441\%     \\
Max Rel. Dif. (\%)  & 0.0578\%     & 0.1666\%     & 0.1034\%     \\
PINN Energy Var. (\%)    & 0.0472\%     & 0.1521\%     & 0.1117\%     \\
FDTD Energy Var. (\%)    & 0.0139\%     & 0.0139\%     & 0.0139\%     \\
\bottomrule
\end{tabular}
\end{table}

This ablation study demonstrates two key findings. First, cumulative energy drift in time-windowed PINN training is an independent failure mode not directly tied to field accuracy. Second, physics-based energy constraints must be designed with care: the global Poynting formulation not only fails to suppress drift but actively worsens it relative to having no constraint, whereas the local formulation used in \textbf{pinnA} provides effective energy stabilization. The critical distinction lies in whether the constraint operates at the pointwise or integral level, a design choice that determines whether local energy violations are corrected where they occur or allowed to spread and compound through the cancellation-of-errors mechanism.
\newpage

\section{Conclusion}
Maxwell's equations constitute the theoretical foundation of the field of electromagnetism and have been successfully solved for decades using well-established numerical methods such as FDTD, FEM, and MoM. For a new methodology to gain acceptance in such a mature field, it must provide concrete advantages over existing methods, not merely offer an alternative. Although Physics-Informed Neural Networks possess potential advantages such as their mesh-free structure and applicability to inverse problems, a systematic hybrid approach is required for them to compete with classical solvers in terms of accuracy and energy consistency. This study has presented such a hybrid methodology in detail for TM$_z$ mode wave propagation in a 2D PEC cavity and demonstrated that PINN-based electromagnetic solvers can achieve FDTD-level performance.

The developed model achieved field accuracy close to the FDTD reference, with an average NRMSE of 0.09\% and a relative $L^2$ error of 1.01\%. In energy conservation metrics, which constitute a more critical measure of success, average relative error of 0.024\% and maximum mismatch of 0.058\% were attained. These results were obtained without using any labeled data during training, solely through physics-based residual losses. In the lossy medium scenario, the model tracked the expected dissipation curve with a relative error below 0.2\%, demonstrating the generalizability of the approach to different physical regimes.

The central argument of this study is that PINNs have distinct requirements at the theoretical formulation and implementation stages. At the theoretical level, Maxwell's equations implicitly guarantee energy conservation through Poynting's theorem, causality through their hyperbolic structure, and interface continuity through field continuity. However, these guarantees are not automatically satisfied within the PINN framework. The neural network may converge to solutions that violate these physical constraints while minimizing the PDE residual. Therefore, incorporating these theoretically guaranteed structures as explicit regularizers in the loss function during implementation is essential, particularly for suppressing cumulative energy drift across sequential time windows.

The hybrid approach proposed in this study addresses this requirement through three fundamental components: causality-aware weighting that preserves causality during time-marching, interface continuity loss that ensures field continuity between successive windows, and local Poynting-based regularizer that suppresses cumulative energy drift. The results support that each of these components makes a meaningful contribution to the final performance and that they produce a synergistic effect when used together.

Finally, the parenthesis effect reported in this study reveals a sensitivity specific to PINN implementations: two algebraically equivalent code expressions can measurably affect optimization dynamics and long-term energy behavior due to differences in computational graph topology. This sensitivity, absent in classical numerical methods, must be taken into consideration in the development of PINN-based physics solvers.

\newpage
\section{Appendices}
\subsection*{Appendix A: Well-Posed Problem}

\paragraph{Derivation of the Coupled PDE System}

In electromagnetic wave propagation problems, the general formulation of Maxwell’s equations exhibits a highly complex structure from a computational perspective. Therefore, the vector field problem is first reduced to simpler scalar or reduced vector field problems, so that the learning process of the neural network is optimized. One of these reduction approaches is known as modal decomposition.

Before proceeding to the modal decomposition stage, the required equations will be obtained using Maxwell’s equations. Under the assumption of a linear, isotropic, and homogeneous medium, the equations can be arranged by using the constitutive relations $B = \mu H$ and $D=\varepsilon E$;

\begin{align*} \nabla \times \mathbf{E} &= -\mu \frac{\partial \mathbf{H}}{\partial t} && \text{(Faraday's Law)} \\ \nabla \times \mathbf{H} &= \mathbf{J} + \varepsilon \frac{\partial \mathbf{E}}{\partial t} && \text{(Ampere's Law)} \\ \nabla \cdot \mathbf{E} &= \frac{\rho_v}{\varepsilon} && \text{(Gauss' Law)} \\ \nabla \cdot \mathbf{H} &= 0 && \text{(Gauss' Law for Magnetism)} \\ \end{align*}

\textbf{Conversion of Faraday’s Law into Explicit Form;}

\begin{align*}\nabla \times \mathbf{E} &= -\mu \frac{\partial \mathbf{H}}{\partial t}\end{align*}

\begin{align*}\begin{vmatrix} \hat{x} & \hat{y} & \hat{z} \\ \frac{\partial}{\partial x} & \frac{\partial}{\partial y} & \frac{\partial}{\partial z} \\ E_x & E_y & E_z \end{vmatrix} = -\mu \frac{\partial H}{\partial t} \begin{bmatrix} H_x \\ H_y \\ H_z \end{bmatrix}\end{align*}

\begin{align*}\nabla \times E &= \nonumber \\
&\hat{x} (\frac{\partial E_z}{\partial y}-\frac{\partial E_y}{\partial z}) + \hat{y} (\frac{\partial E_x}{\partial z} - \frac{\partial E_z}{\partial x}) + \hat{z} (\frac{\partial E_y}{\partial x} - \frac{\partial E_x}{\partial y})\end{align*}

\begin{align*}
&\hat{x} \left( \frac{\partial E_z}{\partial y} - \frac{\partial E_y}{\partial z} \right) \nonumber 
+\hat{y} \left( \frac{\partial E_x}{\partial z} - \frac{\partial E_z}{\partial x} \right) \nonumber 
+ \hat{z} \left( \frac{\partial E_y}{\partial x} - \frac{\partial E_x}{\partial y} \right) \nonumber \\
&= -\mu \left( \hat{x}\frac{\partial \mathbf{H}_x}{\partial t}
+ \hat{y}\frac{\partial \mathbf{H}_y}{\partial t}
+ \hat{z}\frac{\partial \mathbf{H}_z}{\partial t} \right)
\end{align*}

\textbf{Simplification of Ampere’s Law for a Source-Free Region:}

\begin{align*}\nabla \times \mathbf{H} &= \mathbf{J} + \varepsilon \frac{\partial \mathbf{E}}{\partial t}\end{align*}

If the assumption $\mathbf{J} = \mathbf{0}$ is made inside the cavity resonator;

\begin{align*}\nabla \times \mathbf{H} &= \varepsilon \frac{\partial \mathbf{E}}{\partial t}\end{align*}

= \begin{align*}\begin{vmatrix} \hat{x} & \hat{y} & \hat{z} \\ \frac{\partial}{\partial x} & \frac{\partial}{\partial y} & \frac{\partial}{\partial z} \\ H_x & H_y & H_z \end{vmatrix} = \varepsilon \frac{\partial E}{\partial t} \begin{bmatrix} E_x \\ E_y \\ E_z \end{bmatrix}\end{align*}

= \begin{align*}
\hat{x} (\frac{\partial H_z}{\partial y}-\frac{\partial H_y}{\partial z}) + \hat{y} (\frac{\partial H_x}{\partial z} - \frac{\partial H_z}{\partial x}) + \hat{z} (\frac{\partial H_y}{\partial x} - \frac{\partial H_x}{\partial y})
\end{align*}

\begin{align*}
&\hat{x} \left(\frac{\partial H_z}{\partial y}-\frac{\partial H_y}{\partial z}\right) \nonumber 
+\hat{y} \left(\frac{\partial H_x}{\partial z} - \frac{\partial H_z}{\partial x}\right) \nonumber 
+\hat{z} \left(\frac{\partial H_y}{\partial x} - \frac{\partial H_x}{\partial y}\right)  \nonumber \\
&=\varepsilon \left(\hat{x}\frac{\partial \mathbf{E}_x}{\partial t} 
+ \hat{y}\frac{\partial \mathbf{E}_y}{\partial t} 
+ \hat{z}\frac{\partial \mathbf{E}_z}{\partial t})\right)
\end{align*}

\paragraph{Modal Decomposition Stage}

Modal decomposition is the process of separating electromagnetic fields into transverse and longitudinal components with respect to the propagation direction. With the assumption of propagation in the $z$ direction, in geometries such as a rectangular waveguide or a cavity resonator, electromagnetic fields can be decomposed into Transverse Electric (TE) and Transverse Magnetic (TM) modes. This decomposition is not only a mathematical convenience, but also a natural consequence of the physical boundary conditions.

In electromagnetic wave propagation, modal analysis is performed according to the spatial distribution of the field components. Under the $z$-invariant geometry assumption ($\frac{\partial}{\partial z} = 0$), electromagnetic fields can be decomposed into two fundamental modes: Transverse Magnetic (TM) and Transverse Electric (TE) modes. This decomposition enables the PINN architecture to produce solutions more effectively, because the 6-component vector field problem is reduced to a 3-component reduced field problem.

\textbf{Transverse Magnetic (TM) Mode}

In the TM mode, the component of the magnetic field in the propagation direction (the z direction) is zero. In this case, the field configuration is defined as

$E = (0, 0, E_z)$ and $H = (H_x, H_y, 0)$

This field configuration indicates that the electric field has only a longitudinal component, while the magnetic field remains entirely in the transverse plane. This naturally arises in structures such as rectangular waveguides and cavity resonators under perfect electric conductor (PEC) boundary conditions.

Starting from Faraday’s Law:

\begin{align*}\hat{x} \frac{\partial E_z}{\partial y} - \hat{y}\frac{\partial E_z}{\partial x} = -\mu (\hat{x}\frac{\partial \mathbf{H}_x}{\partial t} + \hat{y}\frac{\partial \mathbf{H}_y}{\partial t})\end{align*}
 
\begin{align*}\boxed{\frac{\partial E_z}{\partial y} = -\mu \frac{\partial \mathbf{H}_x}{\partial t}, \:\:\:\:\:\ \frac{\partial E_z}{\partial x} = \mu \frac{\partial \mathbf{H}_y}{\partial t} }\end{align*}

With Ampere’s Law, the following equality is obtained;

\begin{align*}\hat{z} (\frac{\partial H_y}{\partial x} - \frac{\partial H_x}{\partial y}) = \varepsilon \hat{z}\frac{\partial \mathbf{E}_z}{\partial t}\end{align*}

\begin{align*}\boxed{\frac{\partial H_y}{\partial x} - \frac{\partial H_x}{\partial y} = \varepsilon \frac{\partial \mathbf{E}_z}{\partial t}}\end{align*}

Under the assumption $\frac{\partial}{\partial z} = 0$, the coupled PDE system for the TM mode can be defined as follows;

\begin{align*}\left\{\begin{matrix} \frac{\partial H_y}{\partial x} - \frac{\partial H_x}{\partial y} - \varepsilon \frac{\partial E_z}{\partial t} = 0 \\ \frac{\partial E_z}{\partial y} + \mu \frac{\partial H_x}{\partial t} = 0 \\ \frac{\partial E_z}{\partial x} - \mu\frac{\partial H_y}{\partial t} = 0 \end{matrix}\right.\end{align*}

This three-equation system contains preliminary information about the PINN model to be developed. It can be seen that there will be three components $(E_z, H_x, H_y)$ in the output layer of the neural network, that the primary field will be $E_z$ and will satisfy the dominant wave equation, and that $H_x$ and $H_y$ will be coupled as auxiliary fields through the spatial derivatives of $E_z$.

\newpage

\subsection*{Appendix B: Non-dimensionalization}

Non-dimensionalization is a critical preprocessing step in the numerical solution of partial differential equation systems. This approach is especially vital in gradient-based optimization methodologies such as Physics-Informed Neural Networks, in terms of numerical stability and convergence performance. Normalizing the characteristic scales of the problem parameters through dimensional analysis allows the neural network to learn field components with different magnitudes in a balanced manner.

In this section, a systematic non-dimensionalization procedure is applied to the coupled PDE system obtained for the TM mode. In this study, the interior of the cavity is treated as free space $\varepsilon_r = \mu_r = 1$.

\paragraph{Dimensional Analysis and Field Variables}

The coupled PDE system for the TM mode is given in the previous section;
\begin{align*} &\frac{\partial \mathbf{H}_y}{\partial x} - \frac{\partial \mathbf{H}_x}{\partial y} - \varepsilon \frac{\partial \mathbf{E}_z}{\partial t} = 0 \\ &\frac{\partial \mathbf{E}_z}{\partial y} + \mu \frac{\partial \mathbf{H}_x}{\partial t} = 0 \\ &\frac{\partial \mathbf{E}_z}{\partial x} - \mu\frac{\partial \mathbf{H}_y}{\partial t} = 0 \end{align*}
The physical variables in the system and their SI units are defined as follows:

- $E_z$: $z$-variant component of electric field (V/m)
- $H_x$: $x$-variant component of magnetic field (A/m) 
- $H_y$: $y$-variant component of magnetic field (A/m) 
- $x, y$: spatial coordinates (m) 
- $t$: temporal coordinate (s)
- $\varepsilon$: Electric permittivity (F/m)
- $\mu$: Magnetic permeability (H/m)

\paragraph{Characteristic Scales ve Reference Quantities}
The selection of appropriate reference scales for non-dimensionalization should be based on the problem geometry and the characteristic behavior of the physical phenomenon. The spatial scale L is the characteristic dimension of the cavity resonator (typically the cavity width). The temporal scale T is the characteristic period of the electromagnetic wave. The electric field scale $E_0$ is the characteristic amplitude of the initial pulse or excitation. The magnetic field scale $H_0$ is determined based on the principle of electromagnetic energy conservation.

The fundamental relationships among these scales are derived from electromagnetic theory:
\begin{align*} H_0 &= \frac{E_0}{Z_0} \\ Z_0 &= \sqrt{\frac{\mu_0}{\varepsilon_0}} \\ C &= \frac{1}{\sqrt{\mu_0 \varepsilon_0}} \\ L &= C \cdot T \end{align*}
$Z_0$: intrinsic impedance of free space ($\sim$377 $\Omega$)
$C$: speed of light in vacuum

\paragraph{Non-dimensional Variables Definition}
\begin{align*} \bar{x} &= x / L \\ \bar{y} &= y / L \\ \bar{t} &= t / T \\ \bar{E_z} &= E_z / E_0 \\ \bar{H_x} &= H_x / H_0 \\ \bar{H_y} &= H_y / H_0 \end{align*}

\paragraph{Differential Operator Transformation}
Using the chain rule, the transformation of partial derivative operators to non-dimensional coordinates is obtained: 
\begin{align*} \frac{\partial}{\partial x} &= \frac{1}{L} \cdot \frac{\partial}{\partial \bar{x}} \\ \frac{\partial}{\partial y} &= \frac{1}{L} \cdot \frac{\partial}{\partial \bar{y}} \\ \frac{\partial}{\partial t} &= \frac{1}{T} \cdot \frac{\partial}{\partial \bar{t}} \end{align*}

\paragraph{Systematic Non-dimensionalization Procedure}
\begin{align*} \frac{H_0}{L} \cdot \frac{\partial \bar{H_y}}{\partial \bar{x}} - \frac{H_0}{L} \cdot \frac{\partial \bar{H_x}}{\partial \bar{y}} - \frac{\varepsilon \cdot E_0}{T} \cdot \frac{\partial \bar{E_z}}{\partial \bar{t}} &= 0  \\ \frac{E_0}{L} \cdot \frac{\partial \bar{E_z}}{\partial \bar{y}} + \frac{\mu \cdot H_0}{T} \cdot \frac{\partial \bar{H_x}}{\partial \bar{t}} &= 0  \\ \frac{E_0}{L} \cdot \frac{\partial \bar{E_z}}{\partial \bar{x}} - \frac{\mu \cdot H_0}{T} \cdot \frac{\partial \bar{H_y}}{\partial \bar{t}} &= 0  \end{align*}

\paragraph{Dimensional Coefficient Analysis}
The first equation can be simplified by dividing by $\frac{H_0}{L}$. This operation automatically normalizes the first two terms of the equation, which yields the following expression:

\begin{align*} \frac{\partial \bar{H_y}}{\partial \bar{x}} - \frac{\partial \bar{H_x}}{\partial \bar{y}} - \frac{\varepsilon \cdot E_0 \cdot L}{H_0 \cdot T} \cdot \frac{\partial \bar{E_z}}{\partial \bar{t}} = 0 \quad  \end{align*}

With the algebraic manipulations below, it can be seen that the third term is also normalized.  \begin{align*} \frac{\varepsilon \cdot E_0 \cdot L}{H_0 \cdot T} &= \frac{\varepsilon \cdot E_0 \cdot L}{\frac{E_0}{Z_0} \cdot T} \\ &= \frac{\varepsilon \cdot L \cdot Z_0}{T} \\ &= \frac{\varepsilon \cdot L \cdot \sqrt{\frac{\mu_0}{\varepsilon_0}}}{T} \\ &= \frac{\varepsilon_r \cdot \varepsilon_0 \cdot L \cdot \sqrt{\mu_0}}{T \cdot \sqrt{\varepsilon_0}} \\ &= \frac{\varepsilon_r \cdot \sqrt{\varepsilon_0} \cdot L \cdot \sqrt{\mu_0}}{T} \\ &= \frac{\varepsilon_r \cdot L \cdot \sqrt{\mu_0 \varepsilon_0}}{T} \\ &= \frac{L}{C \cdot T} \cdot \varepsilon_r \\ &=\frac{L}{L} = 1  \quad \quad \text{(L=C.T)}\end{align*}

$\varepsilon = \varepsilon_0 \varepsilon_r$ and $\mu = \mu_0 \mu_r$ 
for free space: $\varepsilon_r = 1$, $\mu_r = 1$ 

Thus, equation is transformed into the following form: \begin{equation*} \frac{\partial \bar{H_y}}{\partial \bar{x}} - \frac{\partial \bar{H_x}}{\partial \bar{y}} - \frac{\partial \bar{E_z}}{\partial \bar{t}} = 0  \end{equation*}

\paragraph{Final Non-dimensional System}

When a similar analysis is applied to the other equations using the normalization factor $\frac{E_0}{L}$,
 \begin{align*} \frac{\partial \bar{E_z}}{\partial \bar{y}} + \frac{\partial \bar{H_x}}{\partial \bar{t}} &= 0 \\ \frac{\partial \bar{E_z}}{\partial \bar{x}} - \frac{\partial \bar{H_y}}{\partial \bar{t}} &= 0 \end{align*}
 
 The fully non-dimensionalized coupled PDE system is obtained as follows:

\begin{align*} \frac{\partial \bar{H_y}}{\partial \bar{x}} - \frac{\partial \bar{H_x}}{\partial \bar{y}} - \frac{\partial \bar{E_z}}{\partial \bar{t}} &= 0 \\[0.5em] \frac{\partial \bar{E_z}}{\partial \bar{y}} + \frac{\partial \bar{H_x}}{\partial \bar{t}} &= 0 \\[0.5em] \frac{\partial \bar{E_z}}{\partial \bar{x}} - \frac{\partial \bar{H_y}}{\partial \bar{t}} &= 0 \end{align*}

\newpage

\subsection*{Appendix C: Metrics}
This appendix provides the formal definitions of all accuracy and energy metrics used in the main text, together with the discretization conventions adopted for collocated and Yee-grid evaluations.

\paragraph{Field Accuracy Metrics}

Let $u(x,y,t)$ denote any field component among $\{E_z, H_x, H_y\}$, and let $u^{\mathrm{PINN}}$ and $u^{\mathrm{FDTD}}$ be the corresponding PINN prediction and FDTD reference sampled on the same evaluation grid at a snapshot time $t$. With $N$ spatial samples, the component-wise RMSE is
\begin{align*}
\mathrm{RMSE}_u(t)
= \sqrt{\frac{1}{N}\sum_{i=1}^{N}\Big(u_i^{\mathrm{PINN}}(t)-u_i^{\mathrm{FDTD}}(t)\Big)^2 }
\end{align*}

We define NRMSE by normalizing RMSE with a \emph{global amplitude scale} extracted from the FDTD reference over all snapshots and the full spatial domain:
\begin{align*}
\mathrm{NRMSE}_u(t)
= \frac{\mathrm{RMSE}_u(t)}{U_0^u}\times 100\%,
\\
U_0^u = \max_{t,x,y}\big|u^{\mathrm{FDTD}}(x,y,t)\big|
\end{align*}
This normalization mitigates the well-known denominator sensitivity of purely relative metrics in nodal regions of wave solutions.

To aggregate three components into a single scalar, we use an RMS combination of the component RMSE values:
{\small
\begin{align*}
\mathrm{RMSE}_{\mathrm{total}}(t)
= \sqrt{\frac{\mathrm{RMSE}_{E_z}(t)^2
+ \mathrm{RMSE}_{H_x}(t)^2
+ \mathrm{RMSE}_{H_y}(t)^2}{3}}
\end{align*}
}

The corresponding amplitude scale is
\begin{align*}
U_0^{\mathrm{total}}=
\sqrt{\big(U_0^{E_z}\big)^2+\big(U_0^{H_x}\big)^2+\big(U_0^{H_y}\big)^2}
\end{align*}
and the aggregated NRMSE is
\begin{align*}
\mathrm{NRMSE}_{\mathrm{Total}}(t)
= \frac{\mathrm{RMSE}_{\mathrm{total}}(t)}{U_0^{\mathrm{total}}}\times 100\%
\end{align*}

As a complementary metric, we report the relative $L_2$ error on the combined vector field
$u=(E_z,H_x,H_y)$:
\begin{align*}
L2_{\mathrm{Total}}(t)
= \frac{\left\|u^{\mathrm{PINN}}(t)-u^{\mathrm{FDTD}}(t)\right\|_2}
{\left\|u^{\mathrm{FDTD}}(t)\right\|_2}\times 100\%
\end{align*}
The combined norm is computed as
\begin{align*}
\left\|u(t)\right\|_2^2=
\left\|E_z(t)\right\|_2^2+\left\|H_x(t)\right\|_2^2+\left\|H_y(t)\right\|_2^2
\end{align*}
with each $\|\cdot\|_2$ taken over the same discrete evaluation grid.

\paragraph{Total Energy and Continuous-Grid Energy Metrics}

In non-dimensional form, the electromagnetic energy density for the TM$_z$ state is
\begin{align*}
\mathcal{u}(x,y,t)=\frac{1}{2}\Big(E_z(x,y,t)^2 + H_x(x,y,t)^2 + H_y(x,y,t)^2\Big)
\end{align*}
and the total energy is
\begin{align*}
E(t)=\int_A \mathcal{u}\, dA
=\frac{1}{2}\int_A\Big(E_z^2+H_x^2+H_y^2\Big)\,dA
\end{align*}

\paragraph{Collocated evaluation grid (PINN-side post-processing).}
When fields are sampled on a uniform collocated grid, the integral is approximated via trapezoidal quadrature weights $w_{ij}$:
\begin{align*}
E^{\mathrm{PINN}}(t)=\frac{1}{2}\sum_{i,j}
\Big(E_z^2+H_x^2+H_y^2\Big)_{ij}\, w_{ij}\,\Delta x\,\Delta y
\end{align*}

\paragraph{FDTD-side energy on a shared evaluation grid.}
For fair post-processing comparisons on a collocated grid, the FDTD fields are sampled consistently at the same spatial locations (and synchronized in time if necessary due to leapfrog staggering). When the FDTD update stores $E_z$ and $(H_x,H_y)$ at half-step shifted times, a standard synchronization is to evaluate the electric contribution at integer time $t^n$ via the half-step average
\begin{align*}
E_z(x,y,t^n)\approx \tfrac{1}{2}\Big(E_z(x,y,t^{n+\tfrac{1}{2}})+E_z(x,y,t^{n-\tfrac{1}{2}})\Big)
\end{align*}
so that all terms entering $E(t^n)$ share a consistent time reference.

\paragraph{Energy mismatch (PINN vs FDTD).}
The instantaneous absolute and relative energy errors are defined as
\begin{align*}
\mathrm{Abs.Error}(t)&=\Big|E^{\mathrm{PINN}}(t)-E^{\mathrm{FDTD}}(t)\Big|\\
\mathrm{Rel.Error}(t)&=\frac{\Big|E^{\mathrm{PINN}}(t)-E^{\mathrm{FDTD}}(t)\Big|}
{E^{\mathrm{FDTD}}(t)}\times 100\%
\end{align*}

\paragraph{Energy conservation (per method).}
To quantify internal conservation relative to each method’s own initial energy,
\begin{align*}
\mathrm{Cons.PINN}(t)
&=\frac{E^{\mathrm{PINN}}(t)-E^{\mathrm{PINN}}(0)}{E^{\mathrm{PINN}}(0)}\times 100\%\\
\mathrm{Cons.FDTD}(t)
&=\frac{E^{\mathrm{FDTD}}(t)-E^{\mathrm{FDTD}}(0)}{E^{\mathrm{FDTD}}(0)}\times 100\%
\end{align*}

\paragraph{Time-Window Interface Energy Jump}

For sequential time-window training, let $\mathcal{W}_k$ and $\mathcal{W}_{k+1}$ denote two consecutive windows with interface time $t_i$. The interface energy jump is
\begin{align*}
\mathrm{Jump}_k
=\frac{\Big|E^{\mathrm{window}_{k+1}}(t_i)-E^{\mathrm{window}_k}(t_i)\Big|}
{E^{\mathrm{window}_k}(t_i)}\times 100\%
\end{align*}
where $E^{\mathrm{window}_k}(t)$ is computed from the window-specific PINN fields using the same collocated-grid quadrature rule.

\paragraph{Yee-Grid Discrete Energy Analysis}

The Yee grid defines $E_z$ at cell centers, $H_x$ at $y$-staggered locations, and $H_y$ at $x$-staggered locations. To test consistency with the \emph{native} FDTD discretization, the PINN fields are evaluated on the same Yee locations (either by direct sampling at those coordinates or by interpolation from a denser collocated grid).

The discrete Yee-grid energy functional is computed as
\begin{align*}
E_{\mathrm{yee}}(t)=\frac{1}{2}\left(\sum E_z^2+\sum H_x^2+\sum H_y^2\right)\Delta x\,\Delta y
\end{align*}
where each sum is taken over the corresponding Yee samples of that component.

Over $N_t$ snapshot times, we summarize the PINN--FDTD mismatch via the mean and maximum absolute differences
\begin{align*}
\overline{\Delta E}
=\frac{1}{N_t}\sum_{t}\left|E_{\mathrm{yee}}^{\mathrm{PINN}}(t)-E_{\mathrm{yee}}^{\mathrm{FDTD}}(t)\right|
\\
\Delta E_{\max}
=\max_{t}\left|E_{\mathrm{yee}}^{\mathrm{PINN}}(t)-E_{\mathrm{yee}}^{\mathrm{FDTD}}(t)\right|
\end{align*}
and the corresponding relative forms normalized by the instantaneous FDTD discrete energy:
\begin{align*}
\overline{\mathrm{Rel.Diff}}
=\frac{1}{N_t}\sum_{t}
\frac{\left|E_{\mathrm{yee}}^{\mathrm{PINN}}(t)-E_{\mathrm{yee}}^{\mathrm{FDTD}}(t)\right|}
{E_{\mathrm{yee}}^{\mathrm{FDTD}}(t)}\times 100\%
\\
\mathrm{Rel.Diff}_{\max}
=\max_{t}
\frac{\left|E_{\mathrm{yee}}^{\mathrm{PINN}}(t)-E_{\mathrm{yee}}^{\mathrm{FDTD}}(t)\right|}
{E_{\mathrm{yee}}^{\mathrm{FDTD}}(t)}\times 100\%
\end{align*}

To summarize each method’s energy band over the trajectory, we also report the normalized variation
\begin{align*}
\mathrm{Var}=\frac{E_{\max}-E_{\min}}{E(0)}\times 100\%
\end{align*}
where $E_{\max}$ and $E_{\min}$ are the maximum and minimum values of the corresponding discrete energy curve over time.




\bibliographystyle{unsrt}
\bibliography{sample.bib}


\end{document}